\documentclass{JHEP3}

\usepackage{latexsym,bm,amsmath,amssymb,amsfonts}
  \usepackage{epsfig,graphics,graphicx}
  \usepackage{slashed}
  \usepackage[latin1]{inputenc}
  \usepackage{mathrsfs}
\usepackage{mathcomp}

  \long\def\comment#1{ }

  \newcommand{\beq}{\begin{eqnarray}}
  \newcommand{\eeq}{\end{eqnarray}}
  
 \def\simge{\mathrel{%
   \rlap{\raise 0.511ex \hbox{$>$}}{\lower 0.511ex \hbox{$\sim$}}}}
\def\simle{\mathrel{
   \rlap{\raise 0.511ex \hbox{$<$}}{\lower 0.511ex \hbox{$\sim$}}}}

\vspace{1.5cm}

\title{\rm \LARGE \bf Finite--Temperature Fractional D2--Branes and the Deconfinement Transition in 2+1 Dimensions}

\author{Gregory C. Giecold \\Institut de Physique Th\'eorique,\\
CEA Saclay,\\
F--91191 Gif-sur-Yvette, France\\
  E-mail: \email{gregory.giecold@cea.fr}}

\abstract{The supergravity dual to N regular and M fractional D2--branes on a cone over $\mathbb{CP}^3$ has a naked singularity in the infrared. One can resolve this singularity and obtain a regular fractional D2--brane solution dual to a confining 2+1 dimensional $\mathcal{N} = 1$ supersymmetric field theory. The confining vacuum of this theory is described by the solution of Cvetic, Gibbons, Lu and Pope~\cite{Cvetic:2001ma}. In this paper, we explore the alternative possibility for resolving the singularity -- the creation of a regular horizon. The black--hole solution we find corresponds to the deconfined phase of this dual gauge theory in three dimensions. This solution is derived in perturbation theory in the number of fractional branes. We argue that there is a first--order deconfinement transition. Connections to Chern--Simons matter theories, the ABJM proposal and fractional M2--branes are presented.}

\begin{document}

\section{Introduction}

Since its inception, the AdS/CFT correspondence~\cite{Maldacena:1997re, Gubser:1998bc, Witten:1998qj} and its various extensions have provided valuable information on gauge theories at strong coupling. In this paper we investigate the deconfinement transition of a $2+1$ dimensional gauge theory by constructing a black hole solution in supergravity.

In order to reach closer connection with QCD or condensed--matter gauge theories there exist different techniques to break some amount of the supersymmetry involved in the gauge/gravity dualities~\cite{Itzhaki:1998dd, Kanitscheider:2008kd}. 
Putting for instance a stack of branes at a singularity in the transverse space results in a dual field theory with lower supersymmetry. More generally, singular points in the compactifying space lead to interesting behaviour in the scaling limit. The geometrical identification of symmetries of the corresponding gauge theory and its amount of supersymmetry appears in~\cite{Morrison:1998cs}. The Klebanov--Witten construction is a particularly interesting example arising from placing $N$ D3--branes at a conical singularity~\cite{Klebanov:1998hh}. The base of the cone is the Einstein manifold $T^{1,1} = \frac{\text{SU}(2) \times \text{SU}(2)}{U(1)}$ with topology $S^2 \times S^3$. The cone over $T^{1,1}$, known as the conifold~\cite{Candelas:1989js}, is defined as the locus $\sum_{i=1}^4 z_i^2 = 0$ in $\mathbb{C}^4$. It is a Calabi--Yau manifold with K\"ahler potential $K = \left( \sum_{i=1}^4 \left| z_i \right|^2 \right)^{2/3}$ and as such indeed preserves $1/4$ supersymmetry. The dual four--dimensional field theory is then $\mathcal{N} = 1$ supersymmetric. The matter field content consists of chiral superfields $A_1$, $A_2$ and $B_1$, $B_2$ in the $\mathbf{(N,\bar{N})}$ and $\mathbf{(\bar{N},N)}$ representations, respectively. Each pair forms a doublet under $\text{SU}(2)$. The theory has a superpotential $\mathcal{W} = \frac{\lambda}{2} \epsilon^{ij} \epsilon^{kl} \text{Tr} A_i B_k A_j B_l$ which preserves the $\text{SU}(2) \times \text{SU}(2) \times U(1)$ isometry of the Einstein metric on $T^{1,1}$.

As a rule, for certain cones it is possible to consider fractional branes which are stuck at the apex and wrap some cycle of the base manifold. For example, going back to the conifold and adding $M$ fractional D3--branes to it changes the gauge group to $\text{SU}(N+M) \times \text{SU}(N)$~\cite{Gubser:1998fp, Klebanov:1999rd, Klebanov:2000nc}. Matching the two gauge couplings to the moduli of IIB string theory on this background leads to a non--vanishing NSVZ beta function~\cite{NSVZ, SV} for $\frac{4\pi}{g_1^2} - \frac{4\pi}{g_2^2}$. The M fractional D3--branes indeed are sources of the magnetic RR 3--form flux through the $S^3$ of $T^{1,1}$ and the RR 3-form field strength's Poincar\'e dual is proportional to the NSNS 3--form field strength. The effective number of D3--branes varies logarithmically with the $\text{AdS}$ radius $r$. The gauge theory intepretation is in terms of a cascade of Seiberg dualities~\cite{Intriligator:1995au, Aharony:2000pp, Dymarsky:2005xt}, i.e. $\text{SU}(N+M) \times \text{SU}(N) \rightarrow \text{SU}(N) \times \text{SU}(N-M)$ and so forth.
 
The Klebanov--Tseytlin solution~\cite{Klebanov:2000nc} is well--behaved at large $r$ but exhibits a naked singularity in the infrared. It was shown in~\cite{Klebanov:2000hb} that in order to remove this singularity the conifold could be replaced by its deformation $\sum_{i=1}^4 z_i^2 = \epsilon^2$. This corresponds to blowing--up the $S^3$ of $T^{1,1}$. The resulting theory is confining and the deformation is the geometrical realization of chiral symmetry breaking. The $U(1)_R$ symmetry is broken to $\mathbb{Z}_{2M}$ by instanton effects. For large $M$, however, $\mathbb{Z}_{2M} \sim U(1)$ and this corresponds to acting as $z_i \rightarrow z_i e^{i\theta}$ on the $\mathbb{C}^4$ embedding coordinates. The deformation breaks this action down to $\mathbb{Z}_2$ while preserving $\mathcal{N} = 1$ supersymmetry.   

Another mechanism for removing the singularity of the Klebanov--Tseytlin solution was developed in~\cite{Buchel:2000ch, Buchel:2001gw, Gubser:2001ri}. The idea is that a non--extremal generalization of the KT solution is expected to develop a regular Schwarzschild horizon which will remove the naked singularity. Unlike the Klebanov--Strassler solution, the KT solution preserves the $U(1)$ symmetry of the Einstein metric of $T^{1,1}$. A non--extremal solution breaks supersymmetry but chiral symmetry is restored in this instance. Reference~\cite{Gubser:2001ri} finds a regular black hole solution via a perturbative expansion in the number of fractional D3--branes. This work is suggestive of a critical temperature $T_c$ where the number of ordinary and fractional branes vanishes at the horizon. This corresponds to an expected reduction in the effective number of degrees of freedom of the dual gauge theory at the phase transition. 

It is our purpose to understand this mechanism for a three--dimensional gauge theory. In~\cite{Cvetic:2001ma}, Cvetic, Gibbons, Lu and Pope (CGLP) derive a regular fractional D2--brane solution. The metric appearing in the CGLP solution involves an asymptotically conical $G_2$ manifold. It is an $\mathbb{R}^3$ bundle over $S^4$. This solution has two supercharges and is then dual to an $\mathcal{N} = 1$ supersymmetric gauge theory in three dimensions. At large distance (small $u$ in our subsequent notation), the geometry becomes a cone over the squashed Einstein metric of the three--dimensional complex projective space $\mathbb{CP}^3$. The resolved solution of~\cite{Cvetic:2001ma} is on par with the Klebanov--Strassler deformed conifold solution in that both are regular solutions. The CGLP solution cures the naked IR singularity caused by flux wrapping a shrinking cycle in~\cite{Herzog:2000rz}. The CGLP solution describes the confining phase of a three--dimensional gauge theory. Since the space ends, the warp factor is finite and so is the tension of a string hanging in this background~\cite{Herzog:2002ss}. This is a hallmark of confinement. The spectrum of minimally--coupled scalars is discrete which is another hint of confinement.
 
In this paper, following analogous work~\cite{Buchel:2000ch} for fractional D3--branes and fractional D1--branes~\cite{Herzog:2001ae}, we show how the singularity appearing in the Herzog--Klebanov (HK) solution~\cite{Herzog:2000rz} is shielded by a regular horizon. We start with non--extremal ordinary and fractional D2--branes probing a cone over $\mathbb{CP}^3$ and from there on build a solution with a horizon. On the gauge theory side, this solution describes the deconfined phase of the underlying three--dimensional field theory. We argue that below some critical temperature, the regular black--hole solution that we derive should be replaced, for thermodynamic reasons, by the CGLP solution. The latter corresponds to the confining vacuum of the dual field theory. 

It is not completely clear what are the gauge group and the field content of the gauge theory whose dual supergravity solution we consider in this paper. According to~\cite{Cvetic:2001ma}, the dual asymptotic field theory should be the same as for regular D2--branes, with gauge group of the special unitary type but with a charge determined by the additional fluxes coming from fractional branes. However, the space transverse to the branes is not $\mathbb{R}^7$ but a cone over $\mathbb{CP}^3$. It is then more appropriate to look for an $\mathcal{N} = 1$ field theory in three dimensions with this space as its classical moduli space of vacua. This analysis is carried in~\cite{Loewy:2002hu} for no fractional branes and suggests that the gauge group is $SU(N) \times SU(N)$. The field content consits in an $\mathcal{N} = 1$ vector multiplet and four $\mathcal{N} = 2$ chiral (eight $\mathcal{N} = 1$ scalar) superfields, one pair in the $\mathbf{(N,\bar{N})}$ representation, another in the conjugate. We have more to say on this in the concluding section. In any case, the gauge group is a product of special unitary groups. If field theory results are of any hint, let us see what one might expect for the gauge dual of the supergravity transition.

All gauge theories with a low--temperature confining phase are thought to possess a non--confining high--temperature phase. This is supported by various pieces of evidence such as lattice studies~\cite{Polyakov:1978vu, Susskind:1979up} or perturbative methods~\cite{Gross:1980br}. In~\cite{Svetitsky:1982gs} the nature of the phase transition for lattice gauge theories with various gauge groups and in various dimensions are presented. Those results rely on dynamical arguments and renormalization group methods to connect the critical behaviour of gauge theories with lower--dimensional spin systems\footnote{See~\cite{Svetitsky:1985ye} for a review and further references.}. Of particular interest to us is the case of three--dimensional gauge theories with gauge group $\text{SU}(N)$ where the critical behaviour should be equivalent to that of a two--dimensional $\mathbb{Z}_N$ spin system. For $N = 2,\ 3 \ \text{or} \ 4$ the general arguments of~\cite{Svetitsky:1982gs} could not rule between a first or second--order phase transition. Of more relevance for the purpose of this paper, for $N \geq 4$ \footnote{This constitutes the case of interest for comparison with theories with a supergravity dual, where $N >> 1$ in order to ignore $\alpha'$ stringy corrections.} the transition was predicted to be either first--order or, if continuous for sufficiently strong coupling, of the Kosterlitz--Thouless type. For the latter kind of transition, thermodynamic quantities display essential singularities at the critical temperature. Relatively recent work on the deconfining phase transition for $\text{SU}(N)$ theories in 2+1 dimensions with $N = 4,\ 5, \ 6$ however suggests that the transition is first order for $N \geq 5$~\cite{Liddle:2005qb}. 
 
Whereas there is some debate concerning the nature of the phase transition from lattice QCD calculations, our supergravity solution hints that the transition between the supersymmetric CGLP solution and the black hole solution we obtain consists in a first--order deconfinement transition for a three--dimensional gauge theory at strong coupling. There is little evidence of a Kosterlitz--Thouless transition. As explained in the Conclusion, it would be interesting within this supergravity solution to further study the free energy to check the nature of the transition and numerically determine the critical temperature at which it vanishes.

The remainder of this paper is organized as follows. In the next section we propose an ansatz for the metric and p--form field strengths for ordinary and fractional D2--branes at finite temperature. We then derive in Section 3, from the IIA equations of motion and the Bianchi identities, a system of mixed, non--linear, second--order differential equations for the functions in our ansatz. We remark on a significant difference from an analysis for D3 or D1-branes, which pertains to the squashing function on the fiber to the transverse space. In Section 4, we present three simple solutions to these equations. Two of them will be used in Sections 5 and 6 as UV and IR boundary conditions for the interpolating non--extremal fractional D2--brane solution we construct from perturbation theory in the number of fractional D2--branes. This perturbative solution is the main result of this paper. We conclude with several suggestions for extension. We mention especially a potential connection to fractional M2--branes and the ABJM proposal.

\section{Non--Extremal Generalization of the Fractional D2--Brane Ansatz}

To construct the non--extremal fractional D2--branes we start with explaining the general ansatz for the metric and the IIA field strengths. It is similar to those described in~\cite{Buchel:2000ch, Buchel:2001gw, Gubser:2001ri} and in~\cite{Herzog:2001ae} for obtaining non--extremal generalizations of fractional D3--branes and D--strings, respectively. They involve adding extra warping functions to the metric which preserve the underlying symmetry of the space transverse to the worldvolume of the branes.

The metric of space transverse to the worldvolume of the D2--branes is the squashed $\mathbb{CP}^3$ Einstein metric. There is a reason why we start with $\mathbb{CP}^3$ instead of one of the simpler manifolds with requisite Betti numbers, such as $S^2 \times S^4$. It has to do with the fact that the cone over a manifold of $\mathbb{CP}^3$ topology admits a smooth resolution. This is the resolution which was used to build the regular supersymmetric fractional D2--brane solution of~\cite{Cvetic:2001ma}. If one is to find a transition from the confined phase whose supergravity dual is the CGLP solution to the deconfined phase of the underlying dual field theory, the UV limits of both supergravity solutions must have the same transverse manifold. 

For this reason the dependence of the p--forms appearing on the cycles of the geometry is almost the same in our ansatz as in the singular Herzog--Klebanov solution~\cite{Herzog:2000rz}. There is a notable difference, though : the 3--form field strength considered in~\cite{Herzog:2000rz} is proportional to $d\rho \wedge \omega_2$. Yet, turning on squashing functions from the transverse space metric of our ansatz below prevents the possibility of an harmonic three--cycle of this type. We discuss this issue further in Section 4.2, where we explain how, for an identically vanishing fiber squashing function $w$ appearing in~\eqref{fb metric}, the 3--form and 4--form field strengths are exactly of the HK type. Higher order corrections result in a non--trivial profile of $w$ associated to sub--leading corrections to the Herzog--Klebanov form for the RR and NSNS field strengths in the UV.

The general ansatz for a 10--d Einstein--frame metric consistent with the underlying symmetries of the squashed three--dimensional complex projective space involves four functions $x$, $y$, $z$ and $w$ of a radial coordinate $u$:
\beq\label{metric ansatz}
ds_{10E}^2 = e^{\frac{5}{2} z} \left( e^{-4x} dX_0^2 + e^{2x} dX_i dX_i \right) + e^{-\frac{3}{2} z} ds_7^2
\eeq
where
\beq\label{cone metric}
ds_7^2 = e^{12 y} du^2 + e^{2y} \left(dM_6\right)^2,
\eeq
\beq\label{fb metric}
\left(dM_6\right)^2 = \lambda^2 e^{- 4 w} \left(D\mu^i \right)^2 + e^{2 w} d\Omega_{4}^2
\eeq
with $d\Omega_4^2$ the metric on the unit 4--sphere. The usual and the squashed Einstein metrics on $\mathbb{CP}^3$ correspond to $\lambda^2 = 1$ and $\lambda^2 = 1/2$, respectively. From here on, we work with the second possibility. The coordinates $\mu^i$ on $\mathbb{R}^3$ are subject to $\mu^i \mu^i = 1$, $i = 1, \ 2, \ 3$. Their covariant derivatives are $D\mu^i \equiv d\mu^i + \epsilon_{ijk} A^j \mu^k$, where $A^j$ refer to $\mathfrak{su}(2)$ Yang--Mills instanton potentials : $A^j = A^j_a T^a = i U \partial^j U$, where $T^a$ stand for $\mathfrak{su}(2)$ generators. The special unitary $2 \times 2$ matrix $U$ can be expressed in terms of Pauli matrices as $U = a_4 + i a^j \sigma^j$. The field strength components $J^i = dA^i + \frac{1}{2} \epsilon_{ijk} A^j A^k$ satisfy the algebra of the unit quaternions: $J^i_{\alpha \gamma} J^j_{\gamma \beta} = - \delta_{ij} \delta_{\alpha \beta} + \epsilon_{ijk} J^k_{\alpha \beta}$.
In~\eqref{metric ansatz}, $X_0$ is the Euclidean time and $X_i$ are the longitudinal D2--brane directions. It should be emphasized that although we switch to Euclidean time in~\eqref{cone metric} and for the external components of Einstein's equations below, intermediate calculations are carried out in real--time.

The metric can be brought into a more familiar form: 
\beq\label{metric familiar form}
ds_{10E}^2 = H(\rho)^{-5/8} \left[ A(\rho) dX_0^2 + dX_i dX_i \right] + H(\rho)^{3/8} \left[ \frac{d\rho^2}{B(\rho)} + \rho^2 \left(dM_6\right)^2 \right],
\eeq
with the redefinitions 
\beq\label{redef}
H(\rho) \equiv e^{-4z - \frac{16}{5}x}, \ \ \ \rho \equiv e^{y + \frac{3}{5}x}, \ \ \ A(\rho) \equiv e^{-6x}, \ \ \ B(\rho)^{-1} d\rho^2 \equiv e^{12y+\frac{6}{5}x} du^2.
\eeq
When $w = 0$ and $e^{5 y} = \rho^5 = \frac{1}{5u}$, the transverse seven--dimensional space is the cone over the squashed $\mathbb{CP}^3$. Small $u$ corresponds to large distances (we indeed assume that $A, B$ and $H$ all approach unity as $\rho \rightarrow \infty$). The function $w(\rho)$ squeezes the $S^2$ fiber relative to the base 4--sphere. It does not affect the symmetries of the $\mathbb{CP}^3$ transverse to fractional branes\footnote{While the metric on $\mathbb{CP}^3$ is usually presented as deriving from a $S^7$ Hopf fibration, it can alternatively be written as a twistor space.}. The extremal D2--brane solution and the more general fractional D2--brane solution on the cone over $\mathbb{CP}^3$ have $x = 0$, $w = 0$. Adding a non--trivial $x(u)$ corresponds to going away from extremality. Our aim is to understand how this changes the extremal D2--brane solution.

It should be noted at this point that the $A$ and $B$ from~\eqref{metric familiar form} turn out to be equal at leading order in the solution we derive in Section 6. This matches the accustomed expectation for a black--hole solution. However, $B$ receives corrections in perturbation theory whereas $A$ is not affected. The solution we find is then different from the usual black D--branes, in that $A \neq B$. It still has a regular horizon with a corresponding Hawking temperature and qualifies as a black hole.

As previously explained, the ansatz for the p--form fields is such that in the UV it is of the same form as for extremal fractional D2--branes~\cite{Herzog:2000rz}: 
\beq\label{4-form ansatz}
\tilde{F}_4 = K(u) e^{\frac{15}{2}z - \frac{\Phi}{2}} du \wedge d^3 x + P \ \Omega_4,
\eeq
\beq\label{NSNS2 ansatz}
H_3 = g_s P \ \Omega_3 .
\eeq
Like the other fields, the dilaton is a function of the radial variable $u$ alone. Here, $\Omega_3 = \star_7 \Omega_{4}$ a harmonic three-form, the Hodge dual being defined with respect to the metric on the cone~\eqref{cone metric}, and
\beq\label{Omega 3}
\Omega_3 = f(u) \omega_1 + g(u) \omega_2 + h(u) \omega_3
\eeq
is a combination of three--forms which are invariant under the $\text{SO}(5)$ isometry group of the base manifold and the $\text{SO}(3)$ isometry group of the $\text{S}^2$ fibers~\cite{Gibbons:1989er, Cvetic:2001ma}. Explicitly,
\beq\label{omega 123}
\omega_1 = D \mu_i \wedge J^i , \ \ \  \omega_2 = du \wedge J , \ \ \  \omega_3 = du \wedge X_2 ,
\eeq
where $J \equiv \mu_i J^i$ and $X_2 \equiv \frac{1}{2} \epsilon_{i j k} \mu^i D\mu^j \wedge D\mu^k$. They satisfy
\beq\label{d omega 123}
d\omega_1 = 0, \ \ \ d\omega_2 = - du \wedge \omega_1, \ \ \ d\omega_3 = - du \wedge \omega_1 , 
\eeq
along with
\begin{align}\label{star omega 123}
\star_7 \omega_1 = e^{6 y} \epsilon_{ijk} \mu^i du \wedge D \mu^j \wedge J^k ,  \  \star_7 \omega_2 = \frac{1}{2} e^{-4 y - 4 w} X_2 \wedge J , \  \star_7 \omega_3 = e^{-4 y + 8 w} J \wedge J . \nonumber\\
\end{align}
To guarantee that the NSNS three--form field strength is harmonic one must have
\beq\label{omega 3 (co)closed}
f' = g + h , \ \ \ e^{6 y} f = \frac{1}{4} \left( e^{-4 w - 4 y} g \right)' ,  \ \ \  e^{6 y} f = \frac{1}{2} \left( e^{-4 y + 8 w} h \right)' . 
\eeq
These are equations for a single independent function $f$, once expressions for $y$ and $w$ are obtained:
\begin{align}\label{f g h expr}
& f = \frac{1}{4} e^{-6y} \left( e^{-4w-4y} \left(1 + 1/2 \ e^{-12 w}\right)^{-1} f' \right)' , \nonumber\\
& g =  \left(1 + 1/2 \ e^{-12 w}\right)^{-1} f' , \nonumber\\
& h = \frac{1}{2} e^{-12 w} g .
\end{align}
Note that $H_3 = dB_2$, with
\beq\label{Btwo}
B_2 = g_s P \ \Big[ \left( \int_0^u h(\rho) d \rho \right) X_2 + \left( \int_0^u g(\rho) d \rho \right) J \Big] ,    
\eeq
up to an exact form.

The $M$ fractional D2--branes (D4--branes wrapping a 2--cycle of the space transverse to the ordinary D2--branes) thus correspond to M units of magnetic flux through the four--cycle of the six--manifold with $P \sim g_s^{3/4} M$. This scaling is derived after~\eqref{zec} in the next section. Ordinary D2--branes are charged electrically under $\tilde{F}_4$ and the function $K(u)$ in \eqref{4-form ansatz} corresponds to the number of ordinary and fractional D2--branes at the scale associated to $u$. The equation of motion for $\tilde{F}_4$
\beq\label{F4 eom}
d\star e^{\frac{\Phi}{2}} \tilde{F}_4 = - g_s^{1/2} F_4 \wedge H_3
\eeq
implies 
\beq\label{K f rlt}
K(u) = Q - 8 g_s^{3/2} P^2 f(u) \int_0^u e^{6 y} f(\rho) d \rho .
\eeq
For the purpose of the calculations leading to~\eqref{K f rlt}, the constraint $\mu^i \mu^i = 1$ allows to take $\mu^i = (0,0,1)$ and we consider a consistent choice for the quaternionic K\"ahler forms~\cite{Cvetic:2001ma} :
\beq\label{J quart spec}
J^1_{12} = J^1_{34} = J^2_{13} = J^2_{42} = J^3_{14} = J^3_{23} = -1, \ \ \ i = 1,2,3 . 
\eeq

At this point, we should note that it is generally not consistent to ask for identically vanishing $f$ and non--trivial $g$ and $h$. However, as long as $w \equiv 0$, which will happen at zeroth--order in the perturbative approach of Section 6,~\eqref{omega 3 (co)closed} gives $g = - h = g_0 e^{4 y}$, with non--necessarily vanishing $g_0$. As a result, the equation of motion for $\tilde{F}_4$ yields 
\beq\label{K f rlt 2}
K(u) = Q - 6 g_s^{3/2} P^2 g_0 \int_0^u g(\rho) d \rho ,
\eeq
instead of~\eqref{K f rlt}.

In what follows we use the ansatz \eqref{metric ansatz}--\eqref{fb metric}, \eqref{4-form ansatz} and \eqref{NSNS2 ansatz} to reduce the remaining equations of motion of IIA supergravity to a system of non--linear, coupled ordinary differential equations describing the radial evolution of $x, y, z, w, f$ and $\Phi$.

\section{Derivation of the Equations of Motion}

Six independent scalars appear in the ansatz~\eqref{metric ansatz},~\eqref{4-form ansatz},~\eqref{NSNS2 ansatz} and we will then need a system of as many ordinary differential equations. Einstein's equation provide five independent equations. The one involving $R_{uu}$ stands apart as it provides a zero energy constraint on integration constants. The equation of motion for the dilaton and the one derived from $H_3 = dB_2$ being (co--)closed provide two nontrivial equations. Like \eqref{F4 eom}, they are derived from the bosonic part of the IIA superstring theory action~\cite{Romans:1985tz} in the Einstein frame:
\begin{align}\label{SIIA}
& \mathcal{S}_{IIA} = \mathcal{S}_{NS} + \mathcal{S}_{R} + \mathcal{S}_{CS}, \nonumber\\
& \mathcal{S}_{NS} = \frac{1}{2 \kappa_{10}^2} \int d^{10}x \sqrt{-G} \Big( R - \frac{1}{2} \partial_M \Phi \partial^M \Phi - \frac{1}{2} e^{-\Phi} \left| H_3 \right|^2 \Big), \nonumber\\
& \mathcal{S}_R = - \frac{1}{4 \kappa_{10}^2} \int d^{10}x \sqrt{-G} \Big( e^{\frac{3\Phi}{2}} \left| \tilde{F}_2 \right|^2 + e^{\frac{\Phi}{2}} \left| \tilde{F}_4 \right|^2 \Big), \nonumber\\
& \mathcal{S}_{CS} = - \frac{1}{4 \kappa_{10}^2} \int B_2 \wedge F_4 \wedge F_4
\end{align}
where $F_2 = dA_1$, $H_3 = dB_2$, $\tilde{F}_4 = dA_3 - A_1 \wedge H_3$ and $2 \kappa_{10}^2 \equiv (2 \pi)^7 \alpha'^4 g_s^2$. Let us specify that
\beq
\int d^{d}x \sqrt{-G} \left| F_p \right|^2 \equiv \int d^dx \sqrt{-G} \frac{1}{p!} G^{M_1N_1} ... G^{M_p N_p} F_{M_1 ... M_p} F_{N_1 ... N_P}.
\eeq
Whenever $P \neq 0$, the $H_3$ equation of motion reduces to the first--order differential equation
\beq\label{H3 ODE}
\left(e^{3z - \Phi}\right)' = - e^{\frac{15}{2} z -\frac{\Phi}{2}} K,
\eeq
while the dilaton equation of motion 
\beq\label{Dil eom}
\nabla^2 \Phi = - \frac{g_s}{12} e^{-\Phi} (H_3)_{MNP} (H_3)^{MNP} + \frac{g_s^{3/2}}{96} e^{\frac{\Phi}{2}} (\tilde{F}_4)_{MNPQ} (\tilde{F}_4)^{MNPQ}
\eeq
yields
\begin{align}\label{Dil ODE}
\Phi'' = P^2 \Big( - g_s^3 e^{3z - \Phi} + \frac{g_s^{3/2}}{2} \ e^{\frac{9}{2}z + \frac{\Phi}{2}} \Big) \Big( e^{-4y-4w} g^2 + 2 e^{-4y + 8w} h^2 + 4 e^{6 y} f^2 \Big) - \frac{g_s^{3/2}}{4} e^{\frac{15}{2}z-\frac{\Phi}{2}} K^2. \nonumber\\
\end{align}
Einstein's equations are $R_{MN} = T_{MN}$ where $R_{MN}$ is the Ricci curvature and the energy--momentum tensor for the relevant field content of IIA supergravity is
\begin{align}\label{T tens}
T_{MN} = &\frac{1}{2} \partial_M \Phi \partial_N \Phi + \frac{g_s}{4} e^{-\Phi} \left( H_M^{\ \ PQ} H_{NPQ} - \frac{1}{12} G_{MN} H^{PQR} H_{PQR} \right) \nonumber\\
& + \frac{g_s^{3/2}}{12} e^{\frac{\Phi}{2}} \left( \tilde{F}_M^{\ \ PQR} \tilde{F}_{NPQR} - \frac{3}{32} G_{MN} \tilde{F}^{PQRS} \tilde{F}_{PQRS} \right).
\end{align}
In order to write down these equations in a convenient form, we will work in an orthonormal frame basis:
\begin{align}\label{oframe1}
& \hat{e}^0 = e^{\frac{5}{4} z - 2x} dX_0, \ \ \ \hat{e}^{x_i} = e^{\frac{5}{4}z + x} dX_i, \ \ \ \hat{e}^{u} = e^{-\frac{3}{4}z + 6y} du, \nonumber\\
& \hat{e}^{\mu^i} = \frac{1}{\sqrt{2}} e^{-\frac{3}{4} z + y - 2w} D\mu^i, \ \ \ \hat{e}^{\alpha} = e^{-\frac{3}{4}z + y + w} g^{\alpha}, \ \ \ i = 1,2,3, \ \ \ \alpha = 1,...,4. 
\end{align}
At this stage, it might not yet be obvious that Einstein's equations are diagonal in this basis. Actually, they will turn out to be so, once we use the Gauss--Codazzi equation~\eqref{GaussCodazzi} that we discuss below, to impose the hypersurface condition $\mu^i \mu^i = 1$. The equations corresponding to $R_{\mu^i \mu^i}$ are identical and similarly for the equations corresponding to $R_{\alpha \alpha}$. The strategy will consist in dealing with the other four Einstein's equations separately from $R_{uu}$ at first. Together with the two field strength equations we will derive thence a second order, nonlinear system of ordinary differential equations in the six warping functions. Finally, we will find out that the $R_{uu}$ relation provides a zero energy constraint, as in~\cite{Buchel:2000ch, Buchel:2001gw, Gubser:2001ri, Herzog:2001ae}.

Let us first compute Ricci's tensors in the above orthonormal basis. We list the non--vanishing components of the curvature two--form $\hat{R}_{MN} = d\hat{\omega}_{MN} + \hat{\omega}_{MP} \wedge \hat{\omega}_{PN}$ found by applying Cartan's structure equations $d\hat{e}^M = - \hat{\omega}_{MN} \wedge \hat{e}^N$,  $\hat{\omega}_{MN} = - \hat{\omega}_{NM}$. The Riemann tensor is obtained from $\hat{R}_{MN} = \frac{1}{2} \hat{R}_{MNPQ} \hat{e}^P \wedge \hat{e}^Q$.
\begin{align}\label{R0M}
& \hat{R}_{0u} = \left( 2 x'' - \frac{5}{4}z'' + 2 \left( \frac{5}{4} z' - 2 x' \right) (x' + 3 y' - z') \right) e^{\frac{3}{2}z - 12y} \hat{e}^0 \wedge \hat{e}^u, \nonumber\\
& \hat{R}_{0x_i} = \left( 2 x' - \frac{5}{4}z' \right) \left( x' + \frac{5}{4} z' \right) e^{\frac{3}{2}z - 12y} \hat{e}^0 \wedge \hat{e}^{x_i}, \nonumber\\
& \hat{R}_{0\mu^i} = \left( 2 x' - \frac{5}{4}z' \right) \left( y' - \frac{3}{4} z' - 2 w' \right) e^{\frac{3}{2}z - 12y} \hat{e}^0 \wedge \hat{e}^{\mu^i}, \nonumber\\
& \hat{R}_{0\alpha} = \left( 2 x' - \frac{5}{4}z' \right) \left( y' + w' - \frac{3}{4} z' \right) e^{\frac{3}{2}z - 12y} \hat{e}^0 \wedge \hat{e}^{\alpha},
\end{align}

\begin{align}\label{RxM}
& \hat{R}_{x_i u} = - \left( x'' + \frac{5}{4} z'' + \left(\frac{5}{4} z' + x' \right) (2z' + x' - 6y') \right) e^{\frac{3}{2}z - 12y} \hat{e}^{x_i} \wedge \hat{e}^{u}, \nonumber\\
& \hat{R}_{x_i x_j} = -  \left(\frac{5}{4} z' + x' \right)^2 e^{\frac{3}{2}z - 12y} \hat{e}^{x_i} \wedge \hat{e}^{x_j}, \nonumber\\
& \hat{R}_{x_i \mu^j} =  \left(\frac{5}{4} z' + x' \right) \left( \frac{3}{4} z' + 2w' - y' \right) e^{\frac{3}{2}z - 12y} \hat{e}^{x_i} \wedge \hat{e}^{\mu^j}, \nonumber\\
& \hat{R}_{x_i \alpha} = -  \left(\frac{5}{4} z' + x' \right) \left( y' + w' - \frac{3}{4}z' \right) e^{\frac{3}{2}z - 12y} \hat{e}^{x_i} \wedge \hat{e}^{\alpha},
\end{align}

\begin{align}\label{RmuM}
\hat{R}_{\mu^i u} = & \left( \frac{3}{4} z'' + 2 w'' - y'' - \left(\frac{3}{4}z' + 2 w' - y' \right) \left(5 y' + 2 w' \right) \right) e^{\frac{3}{2}z - 12y} \hat{e}^{\mu^i} \wedge \hat{e}^{u} \nonumber\\ & - \frac{3}{\sqrt{2}} w' e^{- 5y - 2 w} \epsilon_{ijk} J^j \mu^k, \nonumber
\end{align}
\begin{align}
\hat{R}_{\mu^i \mu^j} = - \frac{1}{2} \left( \frac{3}{4} z' + 2 w' - y' \right)^2 e^{\frac{3}{2} z - 12y} \hat{e}^{\mu^i} \wedge \hat{e}^{\mu^j} + \left(\frac{1}{4} e^{-6w} - 1 \right) \epsilon_{ijk} J^k, \nonumber
\end{align}
\begin{align}
\hat{R}_{\mu^i \alpha} = & - \frac{3}{2 \sqrt{2}} w' e^{\frac{3}{2} z - 7 y - 4 w} \epsilon_{ijk} J^j_{\alpha \beta} \mu^k \hat{e}^{u} \wedge \hat{e}^{\beta} + \frac{1}{2} e^{\frac{3}{2} z - 2 y - 2 w} \left( \frac{1}{4} e^{-6w} - 1\right) \epsilon_{ijk} J^k_{\alpha \beta} \hat{e}^{\mu^j} \wedge \hat{e}^{\beta} \nonumber\\
& + \left( \frac{1}{8} e^{\frac{3}{2} z - 2 y - 8 w} + \frac{1}{\sqrt{2}} \left( \frac{3}{4} z' + 2 w' - y' \right) \left(y' + w' - \frac{3}{4}z' \right) e^{\frac{3}{2} z - 12 y} \right) \hat{e}^{\mu^i} \wedge \hat{e}^{\alpha}, \nonumber\\ 
\end{align}

\begin{align}
\hat{R}_{\alpha u} = &\left( \frac{3}{4} z'' - y'' - w'' + \left( y' + w' - \frac{3}{4} z' \right) \left( 5y' - w' \right) \right) e^{\frac{3}{2} z - 12y} \hat{e}^{\alpha} \wedge \hat{e}^u \nonumber\\
&- \frac{3}{2 \sqrt{2}} w' e^{\frac{3}{2} z - 7 y - 4 w}\epsilon_{ijk} J^{j}_{\alpha \beta} \mu^k \hat{e}^{\mu^i} \wedge \hat{e}^{\beta}, \nonumber
\end{align}
\begin{align}\label{Ralphabeta}
\hat{R}_{\alpha \beta} = & R_{\alpha \beta} - \frac{1}{8} e^{\frac{3}{2} z - 2 y - 8w} \left(\delta_{ij} - \mu^i \mu^j \right) \left( J^i_{\alpha \beta} J^j_{\rho \sigma} + J^i_{\alpha \rho} J^j_{\beta \sigma} \right) \hat{e}^{\rho} \wedge \hat{e}^{\sigma} \nonumber\\ 
& - \left(y' + w' - \frac{3}{4}z' \right)^2 e^{\frac{3}{2}z - 12y} \hat{e}^{\alpha} \wedge \hat{e}^{\beta} + \frac{1}{2} \left(\frac{1}{4} e^{-6w} - 1 \right) e^{\frac{3}{2}z - 2 y - 2 w} \epsilon_{ijk} J^k_{\alpha \beta} \hat{e}^{\mu^i} \wedge \hat{e}^{\mu^j} \nonumber\\
& + \frac{3}{\sqrt{2}} w' e^{\frac{3}{2} z - 7 y - 4 w}  \epsilon_{ijk} J^j_{\alpha \beta} \mu^k \hat{e}^u \wedge \hat{e}^{\mu^i}, \nonumber\\
\end{align}
where $R_{\alpha \beta} = 3 e^{\frac{3}{2}z - 2 y - 2 w} \delta_{\alpha \beta}$ is the curvature two--form on the base manifold $S^4$. Some of the components listed above were already derived in~\cite{Gibbons:1989er}.
To calculate the curvature with the hypersurface condition imposed we use the Gauss--Codazzi equation~\cite{MST:1973}
\beq\label{GaussCodazzi}
\hat{R}_{u^2 = 1}^{MNPQ} = \hat{R}_{STUV} h^{MS} h^{NT} h^{PU} h^{QV} + \chi^{MP} \chi^{NQ} - \chi^{MQ} \chi^{NP},
\eeq
where $h_{MN}$ is the orthonormal frame metric on the projected space:
\beq\label{proj metric}
h_{MN} = \delta_{MN} - \mu^M \mu^N,
\eeq
with $\mu^{M} = \left( \mu^0, \mu^{x_i}, \mu^{u}, \mu^{\mu^i}, \mu^{\alpha} \right) = \left( 0, 0, 0, \mu^i, 0 \right)$ the orthonormal frame components of the unit vector orthogonal to the hypersurface. $\chi_{MN} = h_{MP} h_{NQ} \hat{\nabla}_P \mu_Q$ denote the components of the second fundamental form of the hypersurface. The second fundamental form corresponds to the projection of the gradient to the normal of the hypersurface onto this hypersurface. The only non--vanishing components of $h_{MN}$ and $\chi_{MN}$ are 
\begin{align}\label{h and chi}
&h_{00} = 1, \ h_{x_i x_j} = \delta_{i j}, \ h_{u u} = 1, \ h_{\mu^i \mu^j} = \delta_{i j} - \mu^i \mu^j, \ h_{\alpha \beta} = \delta_{\alpha \beta}, \ \chi_{\mu^i \mu^j} = \sqrt{2} e^{\frac{3}{4}z + 2w - y} h_{\mu^i \mu^j}. \nonumber\\
&
\end{align}

From the field strengths, it is straightforward to check that $T_{00} = T_{x_1 x_1}$\footnote{We recall that, once every contribution is computed, we revert to a Euclidean frame.}. However $R_{00} = e^{\frac{3}{2} z -12 y} \left( 2 x'' - \frac{5}{4} z'' \right)$ and $R_{x_{1,2} x_{1,2}} = - e^{\frac{3}{2} z -12 y} \left(x'' + \frac{5}{4} z'' \right)$. The first two non--redundant Einstein's equations then allow us to solve for $x(u)$ exactly:
\beq\label{x exact}
x'' = 0, \ \ \ x = a u, \ \ \ a > 0.
\eeq
The same behaviour was found for the function $x(u)$ in the case of non--extremal fractional D3--branes and for D--strings~\cite{Buchel:2000ch, Buchel:2001gw, Gubser:2001ri, Herzog:2001ae}. The factor $a$ is identified with the degree of non--extremality.
Having solved for $x(u)$, we can use either of $R_{00}$ or $R_{x_i x_i}$ to derive an equation for $z(u)$:
\begin{align}\label{z eom}
20 z'' = P^2 \Big( 4 g_s^3 & e^{3z - \Phi} + 6 g_s^{3/2} e^{\frac{9}{2}z + \frac{\Phi}{2}} \Big) \Big( e^{-4y-4w} g^2 + 2  e^{-4y + 8 w} h^2 + 4  e^{6y} f^2 \Big) \nonumber\\ + & 5 g_s^{3/2} e^{\frac{15}{2}z - \frac{\Phi}{2}} K^2 . 
\end{align}
Note that in the extremal case $x = 0$, $z = - \Phi$ and $h^{-5/8} = e^{5/2 z} = h^{-1/2} e^{-\Phi/2}$. This means that the Einstein frame metric in the extremal case can be obtained from the string frame metric through the standard procedure of multiplying by $e^{-\Phi/2}$.

In order to find the differential equations for $y(u)$ and $w(u)$, we must consider linear combinations of the Einstein's equations involving
\begin{align}\label{Einstein transverse}
& \hat{R}_{\mu^i \mu^j} = \Big( e^{\frac{3}{2}z - 12y} \Big( \frac{3}{4} z'' + 2w'' - y'' \Big) + 2 e^{\frac{3}{2}z - 2 y + 4 w} \Big( 1 + \frac{1}{4} e^{-12w} \Big) \Big) h_{\mu^i \mu^j}, \nonumber\\
& \hat{R}_{\alpha \beta} = \Big( e^{\frac{3}{2}z - 12y} \Big( \frac{3}{4}z'' - y'' - w'' \Big) + e^{\frac{3}{2}z - 2 y - 2 w} \Big( 3 - \frac{1}{2} e^{-6 w} \Big) \Big) \delta_{\alpha \beta}.
\end{align}
Those are easily computed from the curvature two--forms \eqref{R0M}--\eqref{Ralphabeta} and the Gauss--Codazzi condition \eqref{GaussCodazzi}. We set $\Phi_n \equiv \Phi + z$ and $z_n \equiv 15 z - \Phi$. Computing the field strength contribution to Einstein's equations then results in a system of second order differential equations, where we also list below those found previously in \eqref{f g h expr}, \eqref{Dil ODE} and \eqref{z eom}:
%\beq
%\frac{3}{4} z'' - y'' - w'' + e^{10y - 2w} \Big( 3 - \frac{1}{2} e^{-6w} \Big) = \frac{3}{16} g_s^{3/2} e^{\frac{15}{2} z - \frac{\Phi}{2}} K^2 - \frac{1}{16} \Big( 4 g_s^3 e^{3z-\Phi} + 6 g_s^{3/2} e^{\frac{9}{2} z + \frac{\Phi}{2}} P^2 \Big) \Big( e^{-4y - 4 w} g^2 + 2 e^{-4y+8w} h^2 + 4 e^{6y} f^2 \Big) + \frac{1}{2} g_s^3 e^{3z -\Phi} \Big( e^{-4y -4w} g^2 + 4 e^{6y} f^2 \Big) + g_s^{3/2} \frac{P^2}{2} e^{\frac{9}{2} z + \frac{\Phi}{2}} \Big( e^{-4y -4w} g^2 + 2 e^{-4y + 8w} h^2 + 4 e^{6y} f^2 \Big),
%\eeq
%\beq
%\frac{3}{4} z'' - y'' + 2 w'' + 2 e^{10y + 4 w} \Big( 1 + \frac{1}{4} e^{-12w} \Big) = \frac{3}{16} g_s^{3/2} e^{\frac{15}{2} z - \frac{\Phi}{2}} K^2 - \frac{1}{16} \Big( 4 g_s^3 e^{3z-\Phi} + 6 g_s^{3/2} e^{\frac{9}{2} z + \frac{\Phi}{2}} P^2 \Big) \Big( e^{-4y - 4 w} g^2 + 2 e^{-4y+8w} h^2 + 4 e^{6y} f^2 \Big) + 2 g_s^3 e^{3z - \Phi} \Big( e^{-4y+8w} h^2 + e^{6y} f^2 \Big) + g_s^{3/2} P^2 e^{\frac{9}{2}z + \frac{\Phi}{2}} \Big( e^{-4y-4w} g^2 + 2 e^{6y} f^2 \Big),
%\eeq 
\begin{align}\label{R T1}
15 y'' = P^2 \Big(& g_s^3 e^{ 1/4(z_n - 3 \Phi_n) } - g_s^{3/2} e^{1/4(z_n + 3 \Phi_n)}  \Big) \Big( e^{-4y-4w} g^2 + 2 e^{-4y+8w} h^2 - 6 e^{6y} f^2 \Big) \nonumber\\ & + 5 e^{10y} \Big(2 e^{4w} + 6 e^{-2w} - \frac{1}{2} e^{-8w} \Big),
\end{align}
\begin{align}\label{R T2}
6 w'' = P^2 \Big( & - g_s^3 e^{ 1/4(z_n - 3 \Phi_n) } + g_s^{3/2} e^{1/4(z_n + 3 \Phi_n)} \Big) \Big( e^{-4y-4w} g^2 - 4 e^{-4y+8w} h^2 \Big) \nonumber\\ & - 2 e^{10y} \Big( 2 e^{4w} - 3 e^{-2w} + e^{-8w} \Big),
\end{align}
\begin{align}\label{phi n}
5 \Phi_{n}'' = P^2 \Big( - 4 g_s^3 e^{1/4 (z_n-3\Phi_n)} + 4 g_s^{3/2} e^{1/4(z_n + 3 \Phi_n)} \Big) \Big( e^{-4y - 4 w} g^2 + 2 e^{-4y+8w} h^2 + 4 e^{6y} f^2 \Big) , \nonumber\\
\end{align}
\begin{align}\label{z n}
\frac{1}{4} z_{n}'' = P^2 \Big( g_s^3 & e^{1/4(z_n - 3 \Phi_n)} + g_s^{3/2} e^{1/4(z_n + 3 \Phi_n)} \Big) \Big( e^{-4y - 4 w} g^2 + 2 e^{-4y+8w} h^2 + 4 e^{6y} f^2 \Big) \nonumber\\ &+ g_s^{3/2} e^{z_n/2} K^2 ,
\end{align}
\beq\label{R T3}
\left( e^{-4w-4y} \left(1 + 1/2 \ e^{-12 w}\right)^{-1} f' \right)' = 4 e^{6 y} f ,
\eeq
%\beq\label{R T4}
%\left(e^{1/4(z_n - 3 \Phi_n)}\right)' = - e^{z_n / 2} K .
%\eeq
We have left out thus far the $u$ components of Einstein's equation. For our metric ansatz
\beq\label{Ruu}
\hat{R}_{uu} = e^{\frac{3}{2} z - 12y} \Big( \frac{3}{4} z'' - 6 y'' - 12 w'^2 - 6 x'^2 + 30 y'^2 - \frac{15}{2} z'^2 \Big)
\eeq
which, using \eqref{R T1}, \eqref{z n}, \eqref{phi n} and \eqref{x exact}, provides the zero energy constraint
\begin{align}\label{zec}
& 30 (y')^2 - \frac{1}{32} (z_{n}')^2 - \frac{15}{32} (\Phi_{n}')^2 - 12 (w')^2 + P^2 \Big( - g_s^3 e^{1/4(z_n -  3 \Phi_n)} + g_s^{3/2} e^{1/4(z_n + 3 \Phi_n)} \Big) \nonumber\\ &+ \frac{1}{2} g_s^{3/2} K^2 e^{\frac{1}{2} z_n} - 2 e^{10y} \Big( 2 e^{4w} + 6 e^{-2w} - \frac{1}{2} e^{-8w} \Big)  = 6 a^2.
\end{align}
Later it will be important to keep track of the dimensions of the parameters involved in this set of equations. Looking at the form of the metric \eqref{metric ansatz} it is natural that $e^{y}$ and $u^{-1/5}$ should have dimension of length, while $x$, $z$ and $w$ stay dimensionless. Since we have set the 10--d gravitational constant $\kappa_{10}^2/8\pi$ to unity (i.e.~all scales are measured in units of the 10--d Planck scale $L_P \sim \left(g_s \alpha'^2 \right)^{1/4}$), from \eqref{zec} we conclude that $K$ and $Q$ have dimension $(\text{length})^5$. Using~\eqref{4-form ansatz} and~\eqref{phi n}, $P$ scales as $(\text{length})^2$ and $f$ as $(\text{length})^{-1}$. The dependence on the Planck length can be restored by rescaling $Q \rightarrow L_P^5 Q$, $P \rightarrow L_P^2 P$ and so on. Denoting the number of ordinary and fractional D2--branes by $N$ and $M$, this means that $Q \sim g_s^{5/4} \alpha'^{5/2} N$, $P \sim g_s^{3/4} \alpha'^{3/2} M$.
 
\section{Three Simple Solutions}

Aside from the extremal D2--brane solution, there exist three other simple solutions to the set of equations \eqref{R T1}--\eqref{R T3}, \eqref{zec}. First of all, there is the extremal fractional D2--brane solution which was mentioned in Section 2. It is the analog of the Klebanov--Tseytlin solution for fractional D3--branes. The second one is the non--extremal ordinary D2--brane solution.\\ Later on, in Section 6, we will derive a non--extremal fractional D2--brane solution from perturbation theory in the number $P$ of fractional D2--branes. The solution we find interpolates between the extremal fractional D2--brane solution in the UV and the ordinary black D2--brane Horowitz--Strominger--like solution~\cite{Horowitz:1991cd} in the IR.\\ The third solution is the analog of the singular, non--extremal D3--brane solution found by Buchel~\cite{Buchel:2000ch}. This solution is non--extremal but has a naked singularity in the IR. 

\subsection{Singular non--extremal fractional D2--brane}

A natural attempt at finding a non--extremal solution is to first switch off the stretching function $w(u)$. One will find that the solution is singular. So, this motivates the necessity of keeping a non--trivial squashing function for regular solutions. It will also happen that this solution approaches the extremal fractional D2--brane solution of the Herzog--Klebanov type from the next subsection as the non--extremality parameter $a \rightarrow 0$.

Upon examination of the system of differential equations \eqref{R T1}--\eqref{R T3}, \eqref{zec}, we notice that as soon as we impose that $\Phi_n' = p$, with $p$ a constant, equation ~\eqref{phi n} requires that $\Phi_n = 0$, up to a constant related to the string coupling $g_s$. The condition $\Phi_n' = p$ is a natural one to impose. The Herzog--Klebanov singular solution of Section 4.2 is indeed such that $e^{\Phi} = g_s H(\rho)^{1/4}$, i.e. $e^{\Phi_n} = g_s e^{-\frac{4}{5}a u}$ with $a = 0$, from~\eqref{redef} and~\eqref{x exact}. It is worth noting that in a similar analysis for fractional D3--branes or D1--branes~\cite{Buchel:2001gw, Gubser:2001ri, Herzog:2001ae}, $w = 0$ automatically implies $\Phi_n'' = 0$. Alternatively, the analogues of the IR asymptotic conditions which derive from the solution of subsection 4.3 below for fractional D2--branes, lead to a source term for $\Phi_n$ which in turn prevents a non--trivial squashing function $w$.

For fractional D2--branes, however, the two conditions are generically disconnected. A notable exception occurs for $f(u)$ identically vanishing, which is of interest for the perturbative solution we build in Section 6. In this case, by inserting $w(u) = 0$ in~\eqref{R T2} and~\eqref{phi n}, equation~\eqref{R T2} rearranges to $- \frac{5}{4} \Phi_{n}'' = 0$.

Allowing for a non--vanishing function $f(u)$, we enforce $w = 0$ and $\Phi_n = 0$. From \eqref{R T3}, $z_n$ must then satisfy the first order equation  
\beq\label{zn fstordr}
\Big( e^{- \frac{z_n}{4}} \Big)' = \Big( Q - 8 g_s^{3/2} P^2 f(u) \int_0^u e^{6y} f(\rho) d\rho \Big).  
\eeq
Equation \eqref{R T1} gives $y'' = \frac{5}{2} e^{10y}$. From the zero--energy constraint \eqref{zec} 
\beq\label{zec sng nex frct}
y' = - \sqrt{b^2 + \frac{1}{2} e^{10y}}, \ \ \ \ \ 5 b^2 \equiv a^2
\eeq
one of the integration constants of this second--order differential equation for $y$ is already determined. Equation \eqref{zec sng nex frct} integrates to $e^{5y} = \frac{\sqrt{2/5} \ a}{\sinh(\sqrt{5} au)}$, with $a > 0$ without loss of generality. As a consequence, \eqref{R T3} gives a massaged equation for $f(u)$:
\beq\label{f fstordr2}
\left( \sinh \left( \sqrt{5} a u \right)^{4/5} f' \right)' = \frac{12}{5} \frac{a^2}{\sinh \left( \sqrt{5} a u \right)^{6/5}} f ,
\eeq
which in turn, once inserted into \eqref{zn fstordr}, gives an expression for $z_n$.

Consider now the Ricci scalar for the metric ansatz \eqref{metric ansatz}--\eqref{fb metric} specialized to the solution above:
%\begin{align*}
%& R = e^{3/2 z -12y}(3/2 z'' - 12y'' - 12 w'^2 - 6x'^2 + 30 y'^2 - 15/2 z'^2) + 4 e^{3/2 z - 2y} (e^{4w} + 3 e^{-2w} - 1/4 e^{-8w}) \nonumber\\
%\end{align*}
\begin{align}\label{Ricci singular}
R = e^{\frac{11}{32}z_n - 12 y} \Big[ - \frac{3}{32}  g_s^{3/2} e^{\frac{z_n}{4}} K^2 + \frac{3}{4} g_s^{9/4} P^2 \Big( e^{-4y} g^2 + 2 e^{-4y} h^2 + 4 e^{6y} f^2 \Big) \Big]
\nonumber\\
\end{align}
As for the Buchel solution in the case of D3--branes or the singular non--extremal fractional D--string solution~\cite{Herzog:2001ae}, this solution turns out to have a naked singularity in the far infrared, i.e. at large $u$. This is apparent once we write an asymptotic expansion for $f(u)$ and $z_n(u)$ in the infrared region:
\beq\label{f far IR}
f = f_0 \left(1+\frac{4}{5} e^{-2 \sqrt{5} a u}\right) + \mathcal{O}\left( e^{-2 \sqrt{5} a u} \right),
\eeq
from which a development for $z_n$ is found by integrating \eqref{zn fstordr}:
\begin{align}\label{zn far IR}
e^{-\frac{z_n}{4}} = C + Q u - \frac{8 \ 2^{4/5} a^{1/5} f_0^2 g_{s}^{3/2} P^2}{3 \ 5^{1/10}} u - \frac{4 \ 2^{4/5} \ 5^{2/5} f_0^2 g_{s}^{3/2} P^2}{9 a^{4/5}} e^{-\frac{6 a u}{\sqrt{5}}} + \mathcal{O}\left( e^{-\frac{6 a u}{\sqrt{5}}} \right) , \nonumber\\
\end{align}
where $f_0$ and $C$ are constants.

The constant and linear terms in \eqref{f far IR} and \eqref{zn far IR} dominate in the far infrared: $f \sim f_0$ and $e^{-\frac{z_n}{4}} \sim u$. Furthermore, $e^{-y} \sim e^{b u}$. Consequently, the $e^{-y}$ terms will dominate the Ricci scalar which will blow up at large $u$, even if we consider the limit $P \rightarrow 0$ where no fractional D2--branes are present. The $P \rightarrow 0$ limit of this singular non--extremal solution does not correspond to the black D2--brane solution. In Subsection 4.3 we demonstrate how the ordinary non--extremal D2--brane solution is encompassed within the ansatz dealt with in this paper.

We define, as is standard, the horizon (if present at all) via $G_{00} \equiv e^{\frac{5}{32}z_n + \frac{5}{32}pu -4au} = 0$. From the asymptotics of our solutions and the constraint imposed by the zero--energy condition, a horizon can possibly develop only as $u \rightarrow \infty$. This non--BPS solution does not develop an horizon shielding the naked singularity.
We are thus led to conclude that keeping $\Phi_n'(u)$ constant cannot prevent a naked singularity. It is remarkable that in the case at hand it is still consistent to keep $w = 0$ with a non--trivial $\Phi_n$. On the other hand, in the analysis pertaining to D3--branes, a source for $\Phi_n$ implies that a distorting function has to be switched on if non--singular solutions are to be found.  

\subsection{The extremal Herzog--Klebanov fractional D2--brane solution}

We come by the extremal fractional D2--brane solution by taking the limit $a \rightarrow 0$ of the singular non--extremal D2--brane solution described in the previous subsection. Explicitly, this yields
\beq\label{xtr frct d2 0}
\Phi_n \rightarrow 0, \ \ \ e^{5y} \rightarrow \frac{\sqrt{2}}{5u}, \ \ \ f \rightarrow f_0 \ u^{4/5}, \nonumber
\eeq
\beq
e^{-\frac{z_n}{4}} \rightarrow C + Q u - \frac{2^{8/5} 5^{4/5}}{9} g_s^{3/2} f_0^2 P^2 u^{12/5} . 
\eeq
For this solution to be well--behaved in the UV limit $u \rightarrow 0$, we have set to zero the coefficient in $f(u)$ of a mode growing as $u^{-3/5}$. The metric \eqref{metric ansatz} becomes asymptotically flat at large distances. However, as explained in the following subsection, we will rather take $C = 0$ which amounts to focusing on the low--energy dynamics of the gauge theory dual to this supergravity background. This solution also develops a naked infrared singularity at $u_{cr}$ given by $u_{cr}^{7/5} = \frac{9}{2^{8/5} \ 5^{4/5}} \frac{Q}{g_s^{3/2} f_0^2 P^2}$. We denote $\tilde{L}$ the value $u_{cr}^{-1/5}$ at which the singularity occurs for this extremal fractional D2--brane solution. The singularity arises from the flux associated with fractional D2--branes, which is supported on a shrinking 4--cycle. Allowing for a non--trivial $\Phi_n$ and further taking a non--vanishing squashing function $w$, which corresponds to resolving the base of the cone to a non--trivial bundle, should lead to a regular supergravity solution. Actually, as discussed at the beginning of Section 2, this singular solution differs from Herzog--Klebanov solution~\cite{Herzog:2000rz} due to the contribution $f(u) D\mu^i \wedge J^i$ to $H_3$. The other contribution is of the type $du \wedge \omega_2$, where $\omega_2$ is proportional to the fundamental form of $\mathbb{CP}^3$ when $f_0 \rightarrow 0$. In order to recover an asymptotic solution of the HK type, we set $f_0 = 0$ in~\eqref{xtr frct d2 0} and take into account the remark expressed below~\eqref{J quart spec}. Equation~\eqref{xtr frct d2 0} is replaced by
\beq\label{xtr frct d2}
\Phi_n \rightarrow 0, \ \ \ e^{5y} \rightarrow \frac{\sqrt{2}}{5u}, \ \ \ f \rightarrow 0, \ \ \ g = - h \rightarrow g_0 \left( \frac{\sqrt{2}}{5} \right)^{4/5}  u^{-4/5} , \nonumber
\eeq
\beq
e^{-\frac{z_n}{4}} \rightarrow C + Q u - 2^{2/5} 5^{6/5} g_{s}^{3/2} g_{0}^2 P^2 u^{6/5} . 
\eeq
From~\eqref{redef}, it is clear that $H_3$ is proportional to $g_s P \frac{d\rho}{\rho^2} \wedge \omega_2$ as it should for this solution to be of HK--type.
In what follows, we take $C = 0$. The solution is endowed with a naked IR singularity at
\beq\label{ucr L tilde}
u_{cr}^{1/5} = \frac{Q}{2^{2/5} 5^{6/5} g_{s}^{3/2} g_{0}^2 P^2} .
\eeq
From there on, we define $\tilde{L} \equiv u_{cr}^{-1/5}$.

\subsection{The non--extremal ordinary black D2--brane}

In this case we impose that there are no fractional D2--branes: $f = 0$, $g = 0$ and $P = 0$. From knowledge of this solution in~\cite{Horowitz:1991cd}, we also set $w = 0$. The system of differential equations \eqref{R T1}--\eqref{R T3}, \eqref{zec} simplifies to
\beq\label{syst ord}
y'' = \frac{5}{2} e^{10y}, \ \ \ z_n'' = 4 g_s^{3/2} Q^2 e ^{\frac{z_n}{2}}, \ \ \ x'' = 0, \ \ \ \Phi_n'' = 0. 
\eeq
This integrates to
\beq\label{syst ord int}
x' = a, \ \ \  \Phi_{n}' = p, \ \ \ y'^2 = b^2 + \frac{1}{2} e^{10y}, \ \ \ z_{n}'^2 = c^{2} + 16 g_s^{3/2} Q^2 e^{\frac{z_n}{2}}.
\eeq
The zero--energy constraint \eqref{zec} further restrains the integration constants to satisfy $6a^2 - 30 b^2 + \frac{c^2}{32} + \frac{15}{32} p^2 =0$. Integrating one last time, we obtain
\beq\label{y zn nonextr ord}
e^{5y} = \frac{\sqrt{2} b}{\sinh(5bu)}, \ \ \ e^{\frac{z_n}{4}} = \frac{c}{4 g_s^{3/4} Q \sinh\left(\frac{c}{4}(u+k) \right)}.
\eeq
This can be written in the variables \eqref{redef} of the more familiar D2--brane form of the metric \eqref{metric familiar form} as
\beq\label{rho ord d2}
\rho^5 = e^{5y+3x} = 2^{3/2} b \frac{e^{-(5b-3a)u}}{1-e^{-10bu}},
\eeq
\beq\label{A ord d2}
A(u) = e^{-6x} = e^{-6au}
\eeq
and
\beq\label{h ord d2}
H(u) = e^{-\frac{z_n}{4} - \frac{\Phi_n}{4} -\frac{16}{5}x} = 4 g_s^{3/4} \frac{Q}{c} \sinh\left(\frac{c}{4}(u+k) \right) e^{-\frac{16}{5} au - \frac{p}{4} u}.
\eeq
The decoupling limit in use for the AdS/CFT correspondence and its various extensions in 2+1 dimensions~\cite{Itzhaki:1998dd} leads to $k = 0$ so that $H$ vanishes as $u \rightarrow 0$. This decoupling limit removes the asymptotic Minkowski region of curved geometry created by a stack of branes. This way we focus only on the throat--like, near--horizon region, i.e.~on the low--energy dynamics of the gauge theory on this stack of branes in the dual picture.

We take $5b = \frac{1}{4} c = 3a$, $p = - \frac{4}{5} a$, which satisfy the zero--energy constraint. We thus recover the usual non--extremal D2--brane solution as \eqref{rho ord d2}--\eqref{h ord d2} take the form
\beq\label{h a rho ord d2}
H(\rho) = \frac{2^{1/2} g_s^{3/4} Q}{5\rho^5},\ \ \ A(\rho) = B(\rho) = 1 - \frac{6 \sqrt{2} a }{5 \rho^5}.
\eeq
The non--extremal ordinary D2--brane solution is then given by $w = 0$ together with
\begin{align}\label{ord d2 sumup}
\Phi_n = - \frac{4}{5} a u, \ \ \ e^{6x} = e^{6au}, \ \ \ e^{-5y} = \frac{5 \sinh(3au)}{3 \sqrt{2} a}, \ \ \ e^{-\frac{z_n}{4}} = g_s^{3/4} \frac{Q}{3 a} \sinh(3au). \nonumber\\
\end{align}

\section{Asymptotics of the Regular Non--Extremal Fractional D2--Branes}

Given that no analytic solution to the system \eqref{R T1}--\eqref{R T3}, \eqref{zec} of second--order differential equations with the required properties~(i)--(ii) outlined below could be found, one has to content with a solution in perturbation theory or from numerical integration. As a first step, we present below numerical solutions to the differential equations found at first--order in $P^2$. 

Regardless of the method used to tackle the system \eqref{R T1}--\eqref{R T3}, \eqref{zec}, a solution must satisfy two natural conditions in the IR $u \rightarrow \infty$ and the UV $u \rightarrow 0$ regions:

(i) it must be a one--parameter (the non--extremality parameter $x' = a$ or, see below, the Hawking temperature) generalization of the extremal fractional D2--brane solution \eqref{xtr frct d2}. We must thus ensure that it approaches the latter solution in the UV and impose the following boundary conditions at $u \rightarrow 0$:
\begin{align}\label{large dist cond}
& x, w, \Phi_n, f \rightarrow 0, \ \ \ K(u) \rightarrow 3 \ 2^{7/5} 5^{1/5} g_s^{3/2} g_0^2 P^2 \left( \frac{5}{6} \tilde{L}^{-1} - u^{1/5} \right) , \nonumber\\
& e^{5y} \rightarrow \frac{\sqrt{2}}{5u}, \ \ \ e^{-\frac{z_n}{4}} \rightarrow 2^{2/5} 5^{6/5} g_s^{3/2} g_0^2 P^2 \ u \left( \tilde{L}^{-1} - u^{1/5} \right) .
\end{align}
It should be emphasized that $\tilde{L}$ scales as $P^2$ so that the leading term in the $u^{1/5}$ expansion is also of leading order in $P^2$.

(ii) For $P \rightarrow 0$ it must reduce to the black D2--brane solution. The latter has a regular Schwarzschild horizon, which if preserved as fractional D2--branes are added leads to the following infrared constraints as $u \rightarrow \infty$:
\begin{align}\label{small dist cond}
& x = au, \ \ \ w \rightarrow w^{\star}, \ \ \ \Phi_{n} \rightarrow -\frac{4}{5}au + \Phi_{n}^{\star}, \ \ \ K \rightarrow K^{\star}, \nonumber\\
& y \rightarrow - \frac{3}{5}au + y^{\star}, \ \ \ z_n \rightarrow - 12 au + z_n^{\star}.
\end{align}
These asymptotics ensure the existence of a regular Schwarzschild horizon for $u \gg 1$. The constants $y^{\star}$ and $z^{\star}$ encode the information on the Hawking temperature and entropy of this horizon. Indeed, the metric for $u \rightarrow \infty$ in the $U$--$X_0$ plane, where $U \equiv e^{-3au}$ is the natural near--horizon variable, takes the form
\beq\label{metric U X0}
ds^2 = U^2 e^{\frac{5}{32} z_n^{\star} + \frac{5}{32} \Phi_n^{\star}} dX_0^2 + \frac{1}{9a^2} e^{- \frac{3}{32} z_n^{\star} - \frac{3}{32} \Phi_n^{\star} + 12 y^{\star}} dU^2.
\eeq
The Hawking temperature $T_H$ is fixed from the periodicity of the Euclidean time $X_0$ that guarantees there is no conical singularity in the $U$--$X_0$ plane:
\beq\label{T H}
T_H = \frac{3a}{2\pi} e^{\frac{z_n^{\star}}{8} + \frac{\Phi_n^{\star}}{8} - 6 y^{\star}}.
\eeq
In the canonical ensemble where temperature and volume are independent quantities the Hawking temperature \eqref{T H} of the event horizon is identified with the temperature of the dual gauge theory in its deconfined phase.
There are generally different possibilities for the physics of the large u asymptotics. The first one (a) is that the $u \rightarrow \infty$ solution develops a regular horizon as will be the case from \eqref{small dist cond}. (b) It is also possible that $H(\rho)$ in \eqref{metric familiar form} vanishes at some finite $u$ before $u = \infty$. This corresponds to a naked singularity. (c) Still another possibility is that $H(\rho)$ vanishes at $u = \infty$. The singularity and the horizon coincide in this case. The next section deals with the first option where our ansatz should be appropriate at sufficiently high temperature.

The effective D2--brane charge $K(u)e^{-\frac{\Phi}{2}}$ corresponds in the gauge theory dual to an effective number of unconfined colour degrees of freedom. As we integrate towards large $u$ $K(u)e^{-\frac{\Phi}{2}}$ decreases. On the gauge theory side this matches the expected behaviour that as we run the scale of theory towards the infrared the number of colours degree of freedom should decrease. There is an significant difference from the D3--brane case~\cite{Buchel:2000ch, Buchel:2001gw, Gubser:2001ri} due to the dependence of the D2--brane flux on the dilaton. Note however that $e^{-\frac{\Phi}{2}} \sim \frac{e^{\frac{3}{8}au}}{(\sinh(3a(u+k)))^{1/8}}$ is decreasing for all $u \geq 0$. Thus the flux will still decrease, with the proviso that fractional D2--branes add a small enough perturbation on the variation of the dilaton. The string coupling should be written in terms of gauge theory variables as $e^{\Phi} \sim g_{YM}^{5/2} N^{1/4} / \Lambda^{5/4}$~\cite{Itzhaki:1998dd}. We denote $\Lambda$ the energy scale on the gauge theory side. $N \sim K e^{-\frac{\Phi}{2}}$ is the number of ordinary D2--branes at the radial variable corresponding to this scale. Then $e^{\frac{9}{8}\Phi} \sim g_{YM}^{5/2} K^{1/4} / \Lambda^{5/4}$. $K$ should decrease in the IR as we add fractional D2--branes. Therefore with perturbative corrections included $e^{\Phi}$ still decreases as $u \rightarrow \infty$.

We expect that for low--enough temperatures the effective number of D2--branes will reach zero at some finite $u$. Above the critical temperature we expect though that the flux will stabilize at some finite value $K^{\star} e^{-\frac{\Phi^{\star}}{2}}$ for large $u$. This number should vanish in the limit that the temperature reaches its critical value. It would be interesting to study the free energy. It should vanish at the critical temperature, marking the transition where the CGLP confining solution is thermodynamically favoured. This would also make it clear that the transition is first--order, instead of the other possibility in 2+1 dimensions, namely Kosterlitz--Thouless.

\section{Perturbation Theory in P}

To construct the regular non--extremal fractional D2--brane solution, we shall start with the non--extremal ordinary D2--brane solution \eqref{ord d2 sumup}. We recall that it corresponds to no fractional branes, $P = 0$. In the singularity--shielding analysis of~\cite{Gubser:2001ri, Herzog:2001ae} for D3--branes and D1--branes respectively, $Q$ is replaced by $K^{\star}$ from a set of conditions similar to \eqref{small dist cond} in order to match onto the asymptotics for the infrared, near--horizon boundary conditions. 
We will then attempt to find the deformation of this starting solution order by order in $P^2$. Actually the relevant expansion parameter happens to be $\lambda = g_0^2 P^2 K_{\star}^{-1} a^{-2/5}$, with $g_0 = \mathcal{O}(1)$ appearing in the large distance asymptotic condition~\eqref{large dist cond}. This ratio is dimensionless and for perturbation theory to apply, the effective D2--brane charge must thus be large enough. It will also become clear that the solution we will build in perturbation theory will correspond to the dual gauge theory at a temperature above the expected critical temperature. As in~\cite{Gubser:2001ri, Herzog:2001ae}, it turns out that a nice feature of perturbation theory around the extremal ($a=0$) D2--brane background is that already the first--order correction in $P^2$ yields the exact extremal fractional D2--brane solution \eqref{xtr frct d2}. This is strong evidence that an interpolating non--extremal solution that we only managed to approach in perturbation theory indeed exists.

It is convenient~\cite{Gubser:2001ri, Herzog:2001ae} to rescale the fields by relevant powers of $P^2$:
\begin{align}\label{rescaled fields}
& f(u) = P F(u), \ \ \  g(u) = g_0 e^{4 y} + P G(u), \nonumber\\
& \Phi_{n}(u) = - \frac{4}{5}au + P^2 \phi(u), \ \ \ w(u) = P^2 \omega(u), \nonumber\\
& y(u) \rightarrow y(u) + P^2 \xi(u), \ \ \ z_{n}(u) \rightarrow z_{n}(u) + P^2 \eta(u) , 
\end{align}
where $y(u)$ and $z_{n}(u)$ on the right-hand side are the corresponding functions appearing in the non--extremal ordinary D2--brane solution \eqref{ord d2 sumup}, which we transcribe here as well: 
\beq\label{resc fields}
e^{-5y} = \frac{5 \sinh(3au)}{3 \sqrt{2}a}, \ \ \ e^{-\frac{z_n}{4}} = g_s^{3/4} \frac{K_{\star}}{3a} \sinh(3au).
\eeq
The first--order correction to $g(u)$ and $h(u)$ are found from $F(u)$ and~\eqref{f g h expr}. The function $f(u)$ is identically vanishing at zero order in perturbation theory, so that $w = 0$ implies $\Phi_{n}'' = 0$. As noticed in Section 4.1, this way the condition $\Phi_{n}' = p$, a constant, does not have to imposed independently as would be the case for non--vanishing $f$.

We must impose further conditions on the correction functions $F$, $\phi$, $\omega$, $\xi$ and $\eta$ so as to match onto the extremal fractional D2--brane asymptotics \eqref{large dist cond} near $u = 0$:
\beq\label{correct cond}
\phi(0) = \omega(0) = \xi(0) = 0, \ \ \ \eta \rightarrow 4 \tilde{L} u^{1/5}.
\eeq
Furthermore, $F(u)$ must tend to zero in the UV.
In order to avoid excessive cluttering, from now on we set $g_s = 1$ by absorbing it into $P$.
The system of mixed, non--linear second--order differential equations \eqref{R T1}--\eqref{R T3} along with the constraint \eqref{zec} becomes
\beq\label{pert 2ode1}
12 \xi'' + \phi'' - 300 e^{10y} \xi + \mathcal{O}(P^2) = 0,
\eeq
\beq\label{pert 2ode2}
\omega'' + 2 e^{10y} \omega + \frac{5}{24} \phi'' + \mathcal{O}(P^2) = 0,
\eeq
\beq\label{pert 2ode3}
5 \phi'' + 12 g_{0}^2 e^{\frac{z_n}{4} + 4 y} \left( e^{\frac{3}{5} a u} - e^{-\frac{3}{5} a u} \right) + \mathcal{O}(P^2) = 0,
\eeq
\begin{align}\label{pert 2ode4}
\eta'' & - 2 e^{\frac{z_n}{2}} K_{\star}^2 \eta - 12 g_{0}^2 e^{\frac{z_n}{4} + 4 y} \left( e^{\frac{3}{5}au} + e^{-\frac{3}{5}au} \right) \nonumber\\ & + 48 g_{0}^2 K_{\star} e^{\frac{z_n}{2}} \int_{0}^{u} e^{4 y} d\rho + \mathcal{O}(P^2) = 0,
\end{align}
\beq\label{pert 2ode5}
\left( e^{-4y} F' \right)' - 6 e^{6 y} F + \mathcal{O}(P^2) = 0
\eeq
and
\begin{align}\label{pert zec}
60 & y' \xi' - \frac{1}{16} z_{n}' \eta' + \frac{3}{4}a \phi' + \frac{1}{4} e^{\frac{z_n}{2}} K_{\star}^2 \eta - e^{\frac{z_n}{4}} \left( e^{\frac{3}{5}a u} - e^{-\frac{3}{5} a u} \right) \nonumber\\ & - 6 g_{0}^2 K_{\star} e^{\frac{z_n}{2}} \int_{0}^{u} e^{4 y} d\rho - 150 e^{10y} \xi  + \mathcal{O}(P^2) = 0. \nonumber\\
\end{align}

\subsection{First--order solutions for $\Phi_n$ and $f$}

The equation of motion \eqref{pert 2ode3} for the order--$P^2$ correction to the dilaton field reduces to
\beq\label{dil pert ode}
\phi'' = -12 g_{0}^2 a^{9/5} K_{\star}^{-1} 2^{2/5} \left( \frac{6}{5} \right)^{9/5} \left( \frac{e^{-3 a u}}{1 - e^{-6 a u}} \right)^{9/5} \left( e^{\frac{3}{5}a u} - e^{-\frac{3}{5}a u} \right), 
\eeq   
which in turn yields (imposing the large $u$ boundary condition that $\phi' \rightarrow 0$)
\beq\label{dil pert lin}
\phi' = \frac{3^{9/5} 2^{6/5}}{5^{4/5}} g_{0}^2 a^{4/5} K_{\star}^{-1} \left( 1 - \frac{1 - e^{-\frac{24}{5} a u} }{\left( 1 - e^{-6 a u} \right)^{4/5}} \right). 
\eeq
The small u asymptotics for its part requires that $\phi \rightarrow 0$ near $u=0$. Consequently, with $v \equiv 1 - e^{-6au}$,
\begin{align}\label{dil pert final}
\phi = \left( \frac{162}{625} \right)^{1/5} g_{0}^2 a^{-1/5} K_{\star}^{-1} \Big[& 5 \ v^{1/5}\, _2F_1\left[\frac{1}{5},\frac{1}{5};\frac{6}{5};v\right] + (-)^{4/5} \log(1 + (-v)^{1/5}) \nonumber\\ &+ \log(1 - v^{1/5}) - (-)^{3/5} \log(1 - (-)^{2/5} v^{1/5}) \nonumber\\ &+ (-)^{2/5} \log(1 + (-)^{3/5} v^{1/5}) \nonumber\\ & - (-)^{1/5} \log(1 - (-)^{4/5} v^{1/5}) - \log(1-v) \Big] , \nonumber\\
\end{align}
where $\, _2F_1\left[a,b;c;z\right]$ denotes Gauss' hypergeometric function. Despite the weird aspect of this solution, it is real and well--behaved, as the plot of the term in square brackets does attest on Figure 1. The asymptotics of~\eqref{dil pert final} are
\beq\label{dil asym small}
\phi = \left( \frac{162}{625} \right)^{1/5} g_{0}^2 a^{-1/5} K_{\star}^{-1} \left[ v - \frac{2}{3} v^{4/5} \right] + \mathcal{O}(v^2), \ \ \ u \rightarrow 0
\eeq
and
\beq\label{dil asym large}
\phi = \phi^{\star} + \frac{4}{5} \left( \frac{162}{625} \right)^{1/5} g_{0}^2 a^{-1/5} K_{\star}^{-1} (1-v) + \mathcal{O}(1-v)^2 , \ \ \ u \rightarrow \infty,
\eeq
where 
\begin{align}\label{phi etoile 0}
\phi^{\star} = \left( \frac{162}{625} \right)^{1/5} g_{0}^2 a^{-1/5} & K_{\star}^{-1} \Big( \sqrt{2 + \frac{2}{\sqrt{5}}} + \log\Big[ \frac{1}{5} \left( 1 + (-)^{1/5} \right)^{(-)^{4/5}} \nonumber\\ & \left(1 - (-)^{2/5} \right)^{-(-)^{3/5}} \left( 1 + (-)^{3/5} \right)^{(-)^{2/5}} \left( 1 - (-)^{4/5} \right)^{-(-)^{1/5}} \Big] \Big) . \nonumber\\
\end{align}
Upon writing the imaginary exponentials in terms of cosines and sines as $(-)^{\alpha} = \cos \alpha + i \sin \alpha$ and using $\log (z) = \log | z | + i  \text{Arg}(z) $, this simplifies to a manifestly real expression :
\begin{align}\label{phi etoile}
\phi^{\star} = \left( \frac{162}{625} \right)^{1/5} g_{0}^2 a^{-1/5} & K_{\star}^{-1} \Big( \frac{1}{2} \sqrt{5-\frac{2}{\sqrt{5}}} \pi -\frac{1}{2} \sqrt{5} \text{arcoth}\left(\sqrt{5}\right)-\frac{5 \log 5}{4} \Big) .
\end{align}

\begin{figure}[htbp]

\begin{center}
\includegraphics[width=0.6\textwidth]{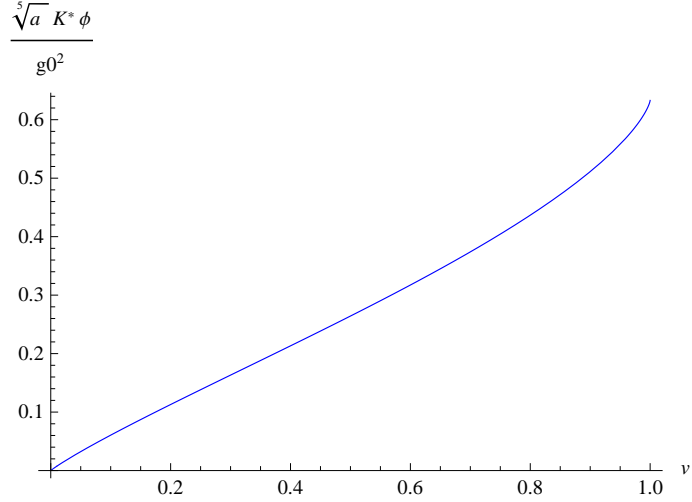}
\caption{\sl Aspect of the first--order solution $\phi (u)$ divided by $g_{0}^2 a^{-1/5} K_{\star}^{-1}$.}
\end{center}
\end{figure}

Switching again to $v = 1 - e^{-6au}$, the differential equation for the first--order solution to the condition that $H_3$ be harmonic~\eqref{pert 2ode5} is 
\beq\label{pert 2ode51}
F'' + \frac{4/5 \ - \ 7/5 \ v}{v(1-v)} F' - \frac{12}{25} \frac{1}{v^2 (1-v)^2} F = 0. 
\eeq
The general solution is of the form
\beq\label{F pert}
F(v) = \frac{ \, _2F_1\left[-\frac{3}{5},-\frac{1}{5},-\frac{2}{5},v\right]}{v^{3/5}} C_1 + v^{4/5} \,  _2F_1\left[\frac{4}{5},\frac{6}{5},\frac{12}{5},v\right] C_2,
\eeq
where $\, _2F_1\left[a,b;c;z\right]$ denotes Gauss' hypergeometric function.
The integration constant $C_1$ and $C_2$ must be such that $F(u)$ increases as degrees of freedom are integrated out down the infrared. This condition yields $C_1 = 0$. The behaviour of $F(v)$ appears on Figure 2.

\begin{figure}[htbp]

\begin{center}
\includegraphics[width=0.6\textwidth]{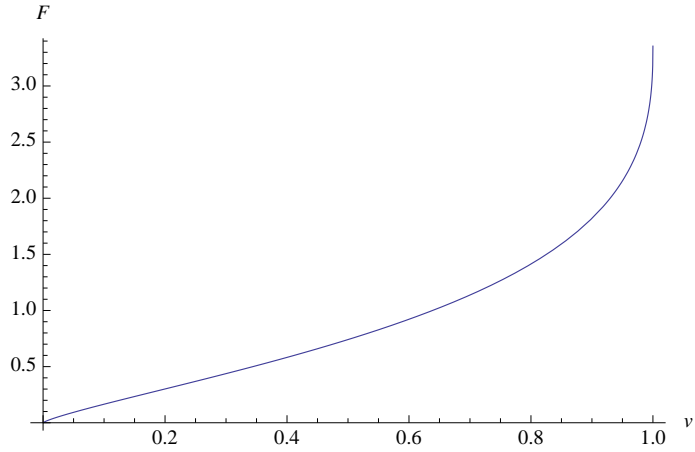}
\caption{\sl Plot of the order--P correction to $f(u)$, with $C_2 = 1$. Even though the conditions needed to derive this solutions are imposed in the IR, it behaves in the UV as the corresponding solution for extremal fractional D2--branes found in Section 4.2.}
\end{center}
\end{figure}

\begin{figure}[htbp]

\begin{center}
\includegraphics[width=0.6\textwidth]{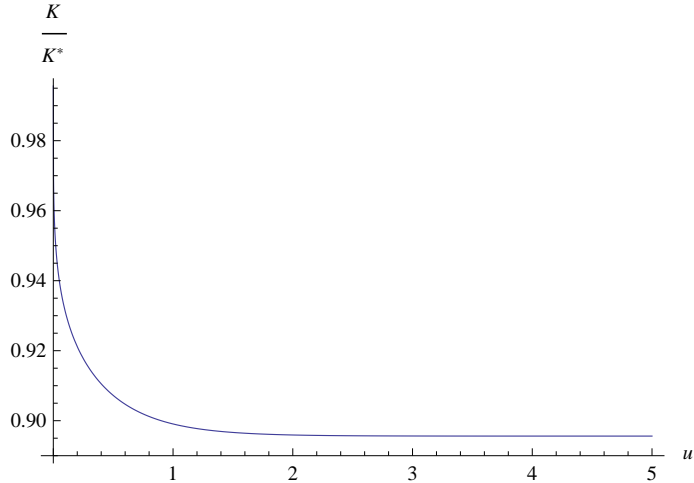}
\caption{\sl The leading--order solution for $\frac{K(u)}{K_{\star}}$ computed from perturbation theory in $P^2$. Here, we put $\lambda = 0.01$. As expected, the effective number of degrees of freedom decreases as we integrate them out flowing to the IR, $v \rightarrow 1$.}
\end{center}
\end{figure}

By way of~\eqref{f g h expr} and~\eqref{K f rlt}, the first--order contribution to $f(u)$ will give rise to a fourth--order correction to the effective number of degrees of freedom.
Figure 3 displays the generic shape of the effective number of ordinary and fractional D2--branes up to second order in $P$ :
\begin{align}\label{K perturb} 
& K(v) = K_{\star} \Big[ 1 - \frac{6}{2^{3/5} 3^{1/5} 5^{4/5}} g_{0}^2 a^{-1/5} K_{\star}^{-1} P^2 (-)^{9/10} \nonumber\\ & \Big( - \frac{\sqrt{\pi} \Gamma(1/10)}{\Gamma(3/5)} + \Big[(1-v)^{1/2} + (1-v)^{-1/2} \Big] \, _2F_1\left[\frac{1}{2},\frac{9}{10};\frac{3}{2};\frac{4}{((1-v)^{1/2} + (1-v)^{-1/2})}\right] \Big) \Big] . \nonumber\\
\end{align} 
We see that, as advertised, the appropriate expansion parameter is $\lambda = g_0^2 a^{-1/5} P^2 K_{\star}^{-1}$. All boundary conditions on $F(u)$ are imposed at $u \rightarrow \infty$ but its $u \rightarrow 0$ limit has a surprise in store. Expanding \eqref{F pert} for small $u$
\beq\label{F pert small}
F = f_0 u^{4/5} + \mathcal{O}(u^{8/5}),
\eeq
identifying $C_2$ with $f_0$, matches, up to $1/\text{length}$ corrections, the exact extremal fractional D2--brane solution \eqref{xtr frct d2 0}. This is strong hint that an exact interpolating solution does exist. We did not manage to find such an analytic solution and have to resort to pertubation theory. Equation \eqref{K perturb} suggests that this approach is reliable as long as $g_0^2 P^2 K_{\star}^{-1} a^{-2/5} << 1$. As already mentioned, this implies the Hawking temperature is high. Expressed in terms of $a$ and $K^{\star}$, \eqref{T H} gives
\beq\label{T H K}
T_H \sim a^{3/10} K_{\star}^{-1/2}.
\eeq 
The critical temperature $T_c$ corresponds to situation where $K^{\star}$ vanishes. This means $a^{2/5} \sim \tilde{L}^7$ at the critical regime. The perturbative approach presented in this paper is valid as long as $T \gg T_c$.

\subsection{Solutions for the other fields}

To determine the correction to $y$, $z_n$ and $w$, the Lagrange method of variation of parameters is particularly suited to solve the linear second--order differential equations~\eqref{pert 2ode1},~\eqref{pert 2ode2},~\eqref{pert 2ode4}. The final product of this recipe guarantees that a differential equation of the type
\beq\label{lin ode ansatz}
\frac{d^2 \psi}{dx^2} + a(x) \frac{d\psi}{dx} + b(x) \psi = c(x) 
\eeq
admits a general solution
\beq\label{solt Lagrange}
\Psi(x) = - \Psi_1(x) \int dx' \frac{c \Psi_2}{W}(x') + \Psi_2(x) \int dx' \frac{c \Psi_1}{W}(x') + c_1 \Psi_1(x) + c_2 \Psi_2(x)
\eeq 
in terms of two linearly independent solutions $\Psi_1$ and $\Psi_2$ for the corresponding homogeneous equation (i.e. \eqref{lin ode ansatz} with $c(x) = 0$). $c_{1,2}$ are arbitrary constants. $W \equiv \Psi_1 \frac{d\Psi_2}{dx} - \Psi_2 \frac{d\Psi_1}{dx}$ is the Wronskian.

Equations \eqref{pert 2ode1}, \eqref{pert 2ode2} and \eqref{pert 2ode4} can be reshuffled into the form \eqref{lin ode ansatz}:
\beq\label{resh 2ode1}
\xi'' - 72 a^2 \frac{e^{-6au}}{\Big(1- e^{-6au}\Big)^2} \xi = \frac{2^{11/5} 3^{9/5}}{5^{9/5}} g_{0}^2 a^{9/5} K_{\star}^{-1} \left( e^{\frac{3}{5}au} - e^{-\frac{3}{5}au} \right) \left( \frac{e^{-3au}}{1-e^{-6au}} \right)^{9/5} ,\nonumber\\
\eeq
\beq\label{resh 2ode2}
\omega'' + \frac{144}{25} a^2 \frac{e^{-6au}}{\Big( 1 - e^{-6au} \Big)^2} \omega = \frac{2^{6/5} 3^{9/5}}{5^{4/5}} g_{0}^2 a^{9/5} K_{\star}^{-1} \left( e^{\frac{3}{5}au} - e^{-\frac{3}{5}au} \right) \left( \frac{e^{-3au}}{1-e^{-6au}} \right)^{9/5} ,\nonumber\\
\eeq
\begin{align}\label{resh 2ode3}
\eta'' - 72 a^2 & \frac{e^{-6au}}{\Big( 1 - e^{-6au} \Big)^2} \eta  = 144 \frac{2^{1/5} 3^{4/5}}{5^{4/5}} g_{0}^2 a^{9/5} K_{\star}^{-1} \left( e^{\frac{3}{5}au} + e^{-\frac{3}{5}au} \right) \left( \frac{e^{-3au}}{1-e^{-6au}}\right)^{9/5} \nonumber\\ & - 1728 \frac{(-)^{9/10}}{2^{3/5} 3^{1/5} 5^{4/5} \Gamma(3/5)} g_{0}^2 a^{9/5} K_{\star}^{-1} \frac{e^{-6 a u}}{\left( 1 - e^{-6 a u} \right)^2} \Big[ -\sqrt{\pi} \Gamma(1/10) \nonumber\\ & \ \ \ \ \ \ \ \ \ \ \ \ \ \ \ \ \ \ \ \ \ + 2 \Gamma(3/5) \cosh(3 a u)\, _2F_1[\frac{1}{2},\frac{9}{10};\frac{3}{2};\cosh(3au)^2] \Big], 
\end{align}
Using $v = 1 - e^{-6au}$ instead as a variable, these equations take on the general form
\beq\label{genansatz}
\frac{d^2\psi}{dv^2} - \frac{1}{1-v} \frac{d\psi}{dv} - \frac{a}{v^2 (1-v)} \psi = c(v),
\eeq
where $a$ and $c(v)$ denote an arbitrary constant and a function, respectively. A solution to the homogeneous part of this equation is $\psi(v) = v^{\nu} \, _2F_1[\nu,\nu;2\nu;v]$ with $a = \nu(\nu - 1)$. The $a=2$ case is somewhat aside in that the two linearly independent solutions reduce to $\frac{1}{v} - \frac{1}{2}$ and $-2 + \frac{v-2}{v} \ln(1-v)$. Equations \eqref{resh 2ode1}--\eqref{resh 2ode3} indeed now read
\beq\label{resh 1}
\xi'' - \frac{\xi'}{1-v} - 2 \frac{\xi}{v^2 (1-v)} = \frac{1}{30} \left(\frac{3 \ 2^{3/2}}{5}\right)^{4/5} g_{0}^2 a^{-1/5} K_{\star}^{-1} \frac{(1-v)^{-1/10}-(1-v)^{1/10}}{v^{9/5}(1-v)^{11/10}} ,\nonumber\\
\eeq
\beq\label{resh 2}
\omega'' - \frac{\omega'}{1-v} + \frac{4}{25} \frac{\omega}{v^2(1-v)} = \frac{1}{12} \left(\frac{3 \ 2^{3/2}}{5}\right)^{4/5} g_{0}^2 a^{-1/5} K_{\star}^{-1} \frac{\left((1-v)^{-1/10}-(1-v)^{1/10}\right)}{v^{9/5} (1-v)^{11/10}} , \nonumber\\
\eeq
\begin{align}\label{resh 3}
& \eta'' - \frac{\eta'}{1-v} - 2 \frac{\eta}{v^2(1-v)} = 4 \frac{2^{1/5} 3^{4/5}}{5^{4/5}} g_{0}^2 a^{-1/5} K_{\star}^{-1} \frac{(1-v)^{1/10} + (1-v)^{-1/10}}{v^{9/5} (1-v)^{11/10}} \nonumber\\ & - 48 \frac{(-)^{9/10}}{2^{3/5} 3^{1/5} 5^{4/5} \Gamma(3/5)} g_{0}^2 a^{-1/5} K_{\star}^{-1} \times \nonumber\\ & \times \frac{-\sqrt{\pi} \Gamma(1/10) + \Gamma(3/5) \left[(1-v)^{1/2} + (1-v)^{-1/2}\right] \, _2F_1[\frac{1}{2},\frac{9}{10};\frac{3}{2},\frac{4}{((1-v)^{1/2} + (1-v)^{-1/2})^2}]}{v^2 (1-v)}. 
\end{align}
Integration constants for $\xi$ are fixed from the boundary conditions that $\xi = 0$ near $v = 0$ and $\xi'(v)$ is finite or vanishes as $v \rightarrow 1$.
This yields
\begin{align}\label{xi pert solt}
\xi = & -\frac{\left(1525+682 \sqrt{5}\right)^{1/10} \pi }{9 \ 2^{4/5} 3^{1/5} 5^{3/5}} g_{0}^2 a^{-1/5} K_{\star}^{-1} \left(\frac{1}{v}-\frac{1}{2}\right) \nonumber\\ & + \frac{1}{18 \ 3^{1/5} 10^{4/5}} g_{0}^2 a^{-1/5} K_{\star}^{-1} \left(-2 + \frac{v-2}{v}\log(1-v)\right) \nonumber\\ & + \frac{1}{72}\left(\frac{5}{48}\right)^{1/5} g_{0}^2 a^{-1/5} K_{\star}^{-1} \Big[8\frac{1+(1-v)^{4/5}}{v^{4/5}} - \frac{2-v}{v} \Gamma(9/5)\Gamma(1/5) \cos \left( \frac{4 \pi}{5}\right) \nonumber\\ & + (2-v) \frac{(1-v)^{4/5}}{v^{4/5}}\, _2F_1[1,1,\frac{9}{5},1-v] + \frac{2-v}{v^{4/5}}\frac{\Gamma(9/5) \Gamma(-1/5)}{\Gamma(4/5)^2}\, _2F_1[1,\frac{1}{5},\frac{6}{5},v]\Big] \nonumber\\
\end{align}
which is displayed on Figure 4.
The solution for $\eta$ is determined from the boundary conditions that $\eta'(v)$ is finite or zero in the limit $v \rightarrow 1$ and $\eta$ stays finite as well for $v \rightarrow 0$. An analytic solution could not be found. On Figure 5 stands the outcome of numerically integrating~\eqref{resh 3}.

\begin{figure}[htbp]

\begin{center}
\includegraphics[width=0.6\textwidth]{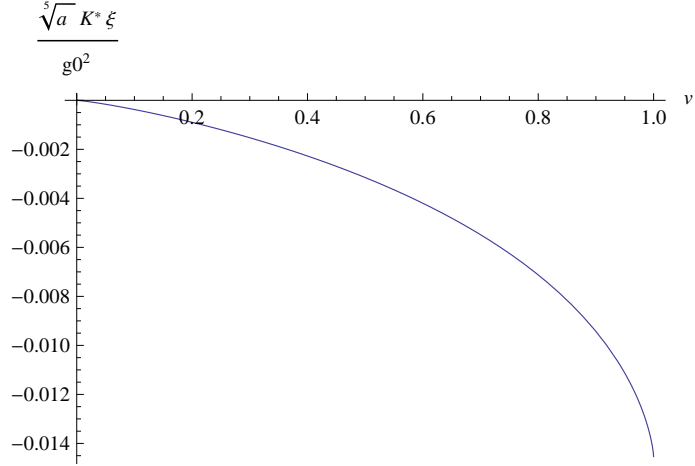}
\caption{\sl First--order correction $\xi(u)$ to $y(u)$ divided by $g_{0}^2 a^{-1/5} K_{\star}^{-1}$ computed from the analytic solution.}
\end{center}
\end{figure}

\begin{figure}[htbp]

\begin{center}
\includegraphics[width=0.6\textwidth]{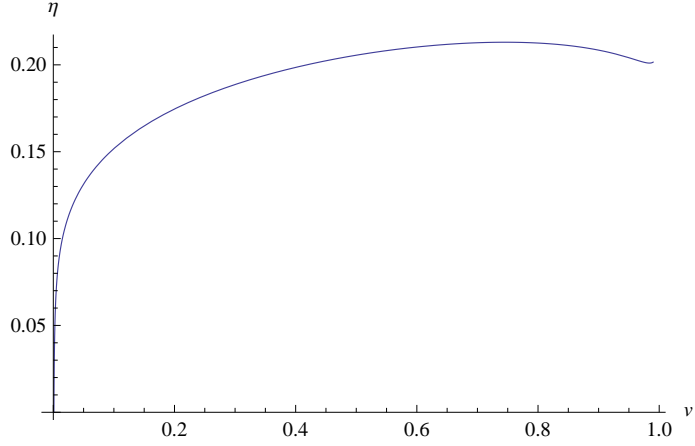}
\caption{\sl First--order correction $\eta(u)$ to $z_{n}(u)$ found numerically by setting $g_{0}^2 a^{-1/5} K_{\star}^{-1} = 0.01$ and enforcing the asymptotic boundary condition discussed in the text.}
\end{center}
\end{figure}

The UV asymptotics $v \rightarrow 0$ of $\xi$ is
\beq\label{xi pert solt large}
\xi = -\frac{g_{0}^2 a^{-1/5} K_{\star}^{-1}}{22 \ 3^{1/5} 10^{4/5}} v^{6/5} +\frac{g_{0}^2 a^{-1/5} K_{\star}^{-1}}{108 \ 3^{1/5} 10^{4/5}} v^2 +\mathcal{O}(v^{11/5}) .
\eeq
The near--horizon $v \rightarrow 1$ expansions of the fields is involved in the computation of the entropy and Hawking temperature of the regular Schwarzschild horizon presented below. For $\xi$ it is
\begin{align}\label{xi pert solt small}
\xi = & \xi^{\star} + \frac{1}{8} \left( \frac{5}{48} \right)^{1/5} g_{0}^2 a^{-1/5} K_{\star}^{-1} (1-v)^{4/5} \nonumber\\ & -\frac{1}{360 \ 3^{1/5} 10^{4/5} \Gamma(4/5)^2} g_{0}^2 a^{-1/5} K_{\star}^{-1} \Big[ 8 \ \Gamma(4/5)^2 \left(-20+\sqrt{10 \left(5+\sqrt{5}\right)} \pi + 5 \log(1-v) \right) \nonumber\\ & + \Gamma(-1/5) \Gamma(9/5) (-1+10 \gamma +10 \log(1-v) + 9 \psi (1/5)+\psi (6/5)) \Big] (1-v) ,\nonumber\\
\end{align}
where
\begin{align}\label{xi star}
\xi^{\star} = \frac{\Gamma(9/5)}{288 \ 3^{1/5} 10^{4/5} \Gamma(4/5)^2} g_{0}^2 a^{-1/5} K_{\star}^{-1} & \Big[ -5 \Big(-40+\sqrt{10 \left(5+\sqrt{5}\right)} \pi \Big) \Gamma(9/5) \nonumber\\ & - 4 \Gamma(-1/5) \Big(\gamma + \psi(1/5) \Big)\Big]
\end{align}
with $\gamma \simeq 0.577216$ being the Euler--Mascheroni constant. The label $\psi(z)$ denotes the digamma function.
The fields $\phi(u)$, $\xi(u)$ and $\eta(u)$ all scale as $g_{0}^2 a^{-1/5} K_{\star}^{-1}$. Since they appear in the perturbative expansion~\eqref{rescaled fields} with a factor of $P^2$ in front, the corrections to $\Phi_n(u)$, $y(u)$ and $z_n(u)$ are all of order $\lambda = P^2 g_{0}^2 a^{-1/5} K_{\star}^{-1} << 1$, the expansion parameter.

Since the $P^2$ corrections to the entropy and the Hawking temperature do not depend on $\omega(u)$ we will just deliver its asymptotics here.
For small $u$ \eqref{pert 2ode2} yields
\beq\label{omega small}
\omega = - \frac{5^{1/5}}{2^{13/5}} g_{0}^2 K_{\star}^{-1} u^{1/5} + \mathcal{O}(u^{4/5}) .
\eeq
The squashing function $\omega$ can be found numerically from the conditions that $\omega(0) = 0$ \eqref{correct cond} and $\omega$ stays finite when $u \rightarrow \infty$. At large $u$ \eqref{pert 2ode2} gives
\beq\label{omega large}
\omega = \omega_{\star} + \frac{5^{6/5}}{2^{24/5} 3^{1/5}} g_{0}^2 a^{-1/5} K_{\star}^{-1} e^{-\frac{24}{5}au} + \mathcal{O}(e^{-5 a u}).
\eeq
Figure 6 shows the behaviour of $\omega(u)$ for $g_{0}^2 a^{-1/5} K_{\star}^{-1} = 0.01$ found from solving numerically~\eqref{resh 2}. 

\begin{figure}[htbp]

\begin{center}
\includegraphics[width=0.6\textwidth]{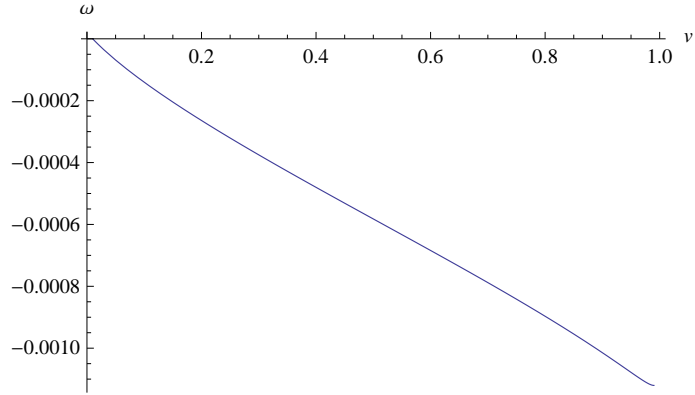}
\caption{\sl First--order correction $\omega(u)$ to the squashing function $w(u)$ found numerically by setting $g_{0}^2 a^{-1/5} K_{\star}^{-1} = 0.01$ and ensuring the asymptotic boundary condition discussed in the text.}
\end{center}
\end{figure}

The corrections to the Hawking temperature and the regular horizon entropy are extracted from the particular form that the metric \eqref{metric ansatz} 
takes when specialized to the perturbative solutions we have just derived. From
\begin{align}\label{mtrc nstz}
ds_{10E}^2 = & \Big( \frac{6a}{K_{\star}} \Big)^{5/8} e^{\frac{5}{32}P^2 (\eta + \phi)} v^{-5/8} \Big( (1-v) dX_0^2 + dX_i dX_i \Big) \nonumber\\ & + (6a)^{1/40} \Big(\frac{2}{25}\Big)^{1/5} K_{\star}^{3/8} v^{-1/40}  \Big[ \frac{1}{6^{1/40}} \frac{2}{25} e^{P^2 \Big(-\frac{3}{32}(\eta + \phi) + 12 \xi \Big)} \frac{dv^2}{v^2(1-v)} \nonumber\\ & + e^{P^2 \Big( -\frac{3}{32}(\eta + \phi) + 2 \xi \Big)}\left( \frac{1}{2} e^{-4 P^2 \omega} \left(D\mu^i \right)^2 + e^{2 P^2 \omega} d\Omega_{4}^2 \right) \Big] , \nonumber\\
\end{align} 
the dependence on the near--horizon asymptotic developments of the entropy density over the temperature squared equals
\beq\label{S T}
\frac{S}{V T^2} = \alpha e^{3P^2 \Big( 6 \xi^{\star} - \frac{\eta^{\star} + \phi^{\star}}{8} \Big)} ,
\eeq 
where $\alpha \equiv 4 \pi^2 \Big( \frac{2}{25} \Big)^{9/5} (6 a^2)^{1/40} K_{\star}^{3/2}$. 
From~\eqref{T H K} and the numerical values of the fields at the horizon, the entropy density over the temperature squared \eqref{S T} tends to a limiting value when the dual deconfined gauge theory is heated up. From the thermodynamic relation $dE = T dS$, the energy of the field theory at strong coupling is about $2/3$ its the free field value, $E \sim \frac{2}{3} T S$. This is to be compared with the celebrated $25 \%$ discrepancy for $\mathcal{N} = 4$ SYM in four dimensions~\cite{Gubser:1996de}. There is an additional contribution from fractional branes to the ratio $\frac{S}{V T^2}$ of the Horowitz--Strominger black D2--brane solution~\cite{Horowitz:1991cd}.

Further numerical work is required. Starting with the UV $u \rightarrow 0$ conditions~\eqref{large dist cond}, the numerical procedure would consist in integrating either the full set of equations or those derived in perturbation theory for the functions in the ansatz. One should then vary the temperature~\eqref{T H} and find the solutions satisfying~\eqref{small dist cond} at a regular horizon shielding the singularity without coinciding with it.

\section{Conclusion}

We  have built to order $P^2$ in the number of fractional branes a regular non--extremal fractional D2--brane perturbative solution. This solution is the supergravity dual of the high--temperature, deconfined phase of the three--dimensional theory whose confined phase supergravity dual was constructed by Cvetic, Gibbons, Lu and Pope~\cite{Cvetic:2001ma}. There are several reasons why one might be interested in this solution and the corresponding field theory. 

The high--temperature limit of QCD in four dimensions may be dominated by the physics of the static theory in three space dimensions, i.e.~Euclidean $\text{QCD}_3$. Within the context of attempting to find an appropriate supergravity dual to QCD, it might then be of interest to start in one lower dimension and see what this can tell about the high--temperature phase. 

Three--dimensional field theories are also of interest for gaining better understanding of the properties of strongly--coupled systems of electrons in condensed matter physics. References~\cite{Hartnoll:2009sz},~\cite{Herzog:2009xv} and~\cite{McGreevy:2009xe} review what supergravity theories have to teach about condensed matter systems belonging to the same universality class of the gauge duals. On this regard, we would like to mention~\cite{Herzog:2009gd} which works out the solution for a baryonic black 3--brane, allowing for a new contribution to the R--R field strengths from which the baryonic $U(1)_B$ gauge field under which their black hole solution is charged stems from upon truncation to five dimensions. 

In~\cite{Aharony:2007vg} a numerical approach was performed for constructing a black hole solution from fractional D3--branes dual to cascading gauge theories. The free energy becomes positive below some critical temperature. This vindicates the suggestion that the supergravity solution is associated to one of the phases separated by a transition which is indeed first--order, as it should from field theory arguments. It would be interesting to carry a similar analysis for the case of the 2+1 dimensional cascading gauge theory dual to fractional D2--branes. This would rule out the possibility of a Kosterlitz--Thouless transition, which the arguments of~\cite{Svetitsky:1982gs} alone cannot prefer over a first--order transition. A Kosterlitz--Thouless phase transition, if present, which would manifest itself through an essential singularity in the free energy.

It would be interesting to understand the spectrum of the field theory dual to the resolved fractional D2--brane solution of~\cite{Cvetic:2001ma} of which we have described the supergravity solution corresponding to the high--temperature, deconfined phase. How this might be guessed from the geometry of the transverse space is described in~\cite{Cvetic:2001zb}. The focus is on the gauge theory dual to the resolved D2--branes and wrapped NS5--branes solution found in~\cite{Cvetic:2001ma}. The base of the cone is $S^3 \times S^3$. Embedding the transverse space with topology $\mathbb{R}^4 \times S^3$ into $\mathbb{R}^8$, one writes the constraint associated to this locus in terms of quaternionic coordinates. Identification of these variables with the matter field of the dual field theory\footnote{The same procedure is used for the conifold~\cite{Klebanov:1998hh}.} provides the representations under which they transform. The most recent attempt to understand the UV regime of the gauge theory we are aware of appears in~\cite{Loewy:2002hu}. This work starts with M2--branes in particular $\text{Spin}(7)$ holonomy backgrounds, $A_8$ and $B_8$, and considers the flow to D2--branes on manifolds of $G_2$ holonomy. A further study of M--theory on the $B_8$ manifold appears in~\cite{Gukov:2001hf}. Reference~\cite{Loewy:2002hu} find that the UV field theory for $N$ D2--branes on the background $\mathcal{M}$, where $\mathcal{M}$ is a cone over $\mathbb{CP}^3$, is $\mathcal{N} = 1$ supersymmetric with $U(N) \times U(N)$ gauge group. The field content fits into an $\mathcal{N} = 1$ vector multiplet and four $\mathcal{N} = 2$ chiral superfields, one pair in the $\mathbf{(N,\bar{N})}$ representation, another in the conjugate. Based on how the gauge groups change from adding fractional branes to the Klebanov--Witten theory, adding $M$ fractional D2--branes might change the gauge group to $U(N) \times U(N+M)$. It would be interesting to verify this in more detail.

Recently, there has been much interest in $\mathcal{N} = 6$ superconformal Chern--Simons matter theories~\cite{Aharony:2008ug}. See~\cite{Klebanov:2009sg} for an excellent review with additional references. When the level $k$ of this $U(N)_{-k} \times U(N)_{k}$ theory describing $N$ coincident M2--branes is such that $k << N <<k^5$, the M--theory circle becomes small and it becomes suitable to describe the gravitational dual using a weakly--curved IIA string theory on a $\text{AdS}_4 \times \mathbb{CP}^3$ geometry orientifolded under the initial M--theory $\mathbb{Z}_k$ projection. A generalization of the ABJM theory to account for the possibility of fractional M2--branes was proposed in~\cite{Aharony:2008gk}. The possibility of duality cascades was further studied in~\cite{Aharony:2009fc}. Further evidence should be provided from dual string theory constructions and this is reported in~\cite{Aharony:2009fc} as work in progress. It might be interesting to build a non--extremal solution from the IIA description of the 't Hooft limit of the ABJM theory. A difference from our work lies in the occurrence of a non--vanishing 2--form field strength in this IIA background. 

More generally, there is a connection between the D2--branes theories we have considered and $\mathcal{N} = 1$ Chern--Simons theories. Indeed, starting from M2--branes on $\text{Spin}(7)$ cones, one can obtain Chern--Simons theories by an orientifolding procedure~\cite{Forcella:2009jj}\footnote{The link between Chern--Simons terms and fluxes in M--theory on a $\text{Spin}(7)$ manifold is discussed in~\cite{Gukov:2001hf}.}. From~\cite{Loewy:2002hu} there is then a flow from C--S theories to D2--branes probing a $G_2$ holonomy manifold. It might also be of interest to study the thermodynamics of those more general flows, with or without fractional branes, which might provide information on the deconfined phase of C-S matter theories. This might hint at how $N^2$ degrees of freedom on D2--branes relate to $N^{3/2}$ on N coincident M2--branes.  

\subsection*{Acknowledgments}
This work was supported in part by a Contrat de Formation par la Recherche of CEA Saclay.
I am grateful to Cl\'ement Ruef, Nick Halmagyi and especially Iosif Bena for comments and discussion on the manuscript. Thanks are due to Cl\'ement Gombeaud for his help with software. I am beholden to the JHEP referee for considerate observations on the preprint.

\end{document}